\documentclass[journal]{IEEEtran}

\usepackage[T1]{fontenc}
\usepackage{amssymb,latexsym,graphicx,times,amsmath,subfigure,threeparttable,booktabs,bm,color,multirow,cite,amsmath,amsfonts,amsthm,pifont}
\usepackage{color}
\usepackage[lined,ruled,commentsnumbered]{algorithm2e}

\makeatletter

\newcommand{\Rmnum}[1]{\expandafter\@slowromancap\romannumeral #1@}
\makeatother

\begin{document}
\title{Alignment-Based Adversarial Training (ABAT)\\ for Improving the Robustness and Accuracy\\ of EEG-Based BCIs}

\author{Xiaoqing~Chen, Ziwei~Wang, and Dongrui~Wu

\thanks{X.~Chen, Z.~Wang and D.~Wu are with the Key Laboratory of the Ministry of Education for Image Processing and Intelligent Control, School of Artificial Intelligence and Automation, Huazhong University of Science and Technology, Wuhan 430074, China. They are also with Shenzhen Huazhong University of Science and Technology Research Institute, Shenzhen, China. Email: \{xqchen, vivi, drwu\}@hust.edu.cn. }}

\markboth{}
{Chen \MakeLowercase{\textit{et al.}}: Alignment-Based Adversarial Training (ABAT) for Improving the Robustness and Accuracy of EEG-Based BCIs}
\maketitle

\begin{abstract}
Machine learning has achieved great success in electroencephalogram (EEG) based brain-computer interfaces (BCIs). Most existing BCI studies focused on improving the decoding accuracy, with only a few considering the adversarial security. Although many adversarial defense approaches have been proposed in other application domains such as computer vision, previous research showed that their direct extensions to BCIs degrade the classification accuracy on benign samples. This phenomenon greatly affects the applicability of adversarial defense approaches to EEG-based BCIs. To mitigate this problem, we propose alignment-based adversarial training (ABAT), which performs EEG data alignment before adversarial training. Data alignment aligns EEG trials from different domains to reduce their distribution discrepancies, and adversarial training further robustifies the classification boundary. The integration of data alignment and adversarial training can make the trained EEG classifiers simultaneously more accurate and more robust. Experiments on five EEG datasets from two different BCI paradigms (motor imagery classification, and event related potential recognition), three convolutional neural network classifiers (EEGNet, ShallowCNN and DeepCNN) and three different experimental settings (offline within-subject cross-block/-session classification, online cross-session classification, and pre-trained classifiers) demonstrated its effectiveness. It is very intriguing that adversarial attacks, which are usually used to damage BCI systems, can be used in ABAT to simultaneously improve the model accuracy and robustness.
\end{abstract}


\IEEEpeerreviewmaketitle

\section{Introduction}

\begin{figure}[htpb]\centering
{\includegraphics[width=0.9\linewidth,clip]{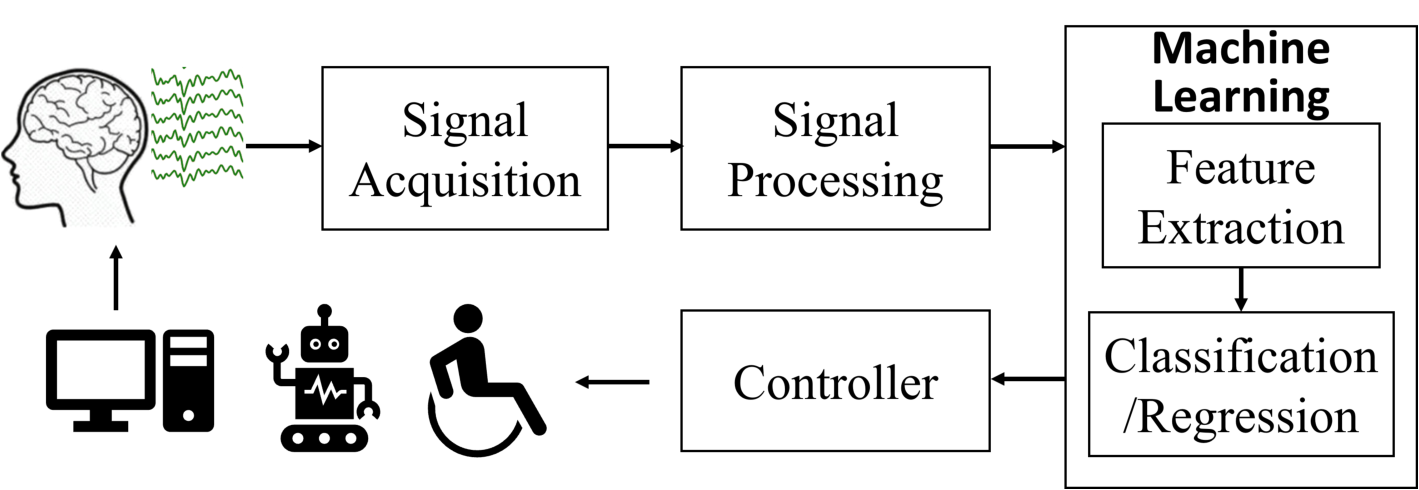}}
\caption{Flowchart of a closed-loop BCI system.} \label{fig:frame}
\end{figure}

\IEEEPARstart{A} brain-computer interface (BCI) establishes a direct communication channel connecting the human brain and a computer \cite{Ienca2018}. Electroencephalogram (EEG), which records the brain's electrical activities from the scalp, is the most commonly utilized input signal in non-invasive BCIs, due to its affordability and ease of use \cite{NicolasAlonso2012}. An EEG-based BCI system typically includes four components: signal acquisition, signal processing, machine learning, and controller, as illustrated in Fig.~\ref{fig:frame}.

\begin{figure}[htpb]\centering
\includegraphics[width=0.6\linewidth,clip]{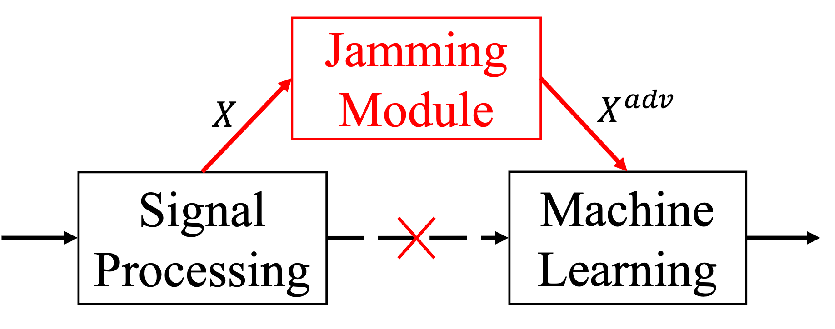}
\caption{The attack framework proposed in \cite{Zhang2019}, which injects a jamming module between signal processing and machine learning to generate adversarial examples.} \label{fig:jamming}
\end{figure}

Most prior research on EEG decoding primarily focused on the accuracy and efficiency of machine learning algorithms \cite{Wu2021}. Nonetheless, a critical discovery by Zhang and Wu \cite{Zhang2019} revealed that adversarial examples, generated using unsupervised fast gradient sign method (FGSM) \cite{Goodfellow2015}, can significantly degrade the performance of deep learning classifiers in EEG-based BCIs. They introduced an attack framework that transforms a benign EEG epoch into an adversarial one by injecting a jamming module before machine learning to add adversarial perturbations, as depicted in Fig.~\ref{fig:jamming}. Furthermore, Zhang \emph{et al.} \cite{Zhang2020} demonstrated that adversarial examples can also fool traditional machine learning classifiers in BCI spellers, misleading them to output an arbitrary (incorrect) character specified by the attacker. Liu \emph{et al.} \cite{Liu2021} and Jung \emph{et al.} \cite{Jung2023} developed approaches to generate universal adversarial perturbations for EEG-based BCIs, making adversarial attacks much easier to implement. Bian \emph{et al.} \cite{Bian2022} employed simple square wave signals to generate adversarial examples, for attacking steady-state visual evoked potential based BCIs. Wang \emph{et al.} \cite{Wang2022} investigated physically constrained adversarial attacks to BCIs. Meng \emph{et al.} \cite{Meng2019} also performed adversarial attacks in EEG-based BCI regression problems.

Adversarial attacks to EEG-based BCIs could have various consequences, from mere user frustration to life-threatening accidents. As pointed out in \cite{Meng2023}, ``\emph{In BCI spellers for Amyotrophic Lateral Sclerosis patients, adversarial attacks may hijack the user's true input and output wrong letters. The user's intention may be manipulated, or the user may feel too frustrated to use the BCI speller, losing his/her only way to communicate with others. In BCI-based driver drowsiness estimation \cite{Wu2016}, adversarial attacks may manipulate the output of the BCI system and increase the risk of accidents. In EEG-based awareness evaluation/detection for disorder of consciousness patients \cite{Li2015}, adversarial attacks may disturb the true responses of the patients and lead to misdiagnosis.}" In military applications, adversarial attacks to BCIs may generate false commands, potentially causing friendly fire \cite{Binnendijk2020}. Consequently, it is very important to develop BCI machine learning models that are robust against adversarial attacks.

Many adversarial defense approaches have been proposed in the literature \cite{Madry2018,Zhang2019a,Zhang2021}, among which robust training \cite{Chen2022} may be the most classical and effective strategy. Adversarial training (AT) \cite{Madry2018} is a representative robust training approach, and many other approaches \cite{Zhang2019a,Zhang2021} can be regarded as its variants. AT solves a minimax problem (the saddle point problem). During training, AT generates adversarial examples along gradients that increase the model's loss to the input, and then minimizes the model's loss on these adversarial examples repeatedly \cite{Madry2018}. This process aims to minimize the model's loss on adversarial examples, but does not explicitly optimize the performance on benign examples. Many studies \cite{Madry2018,Shafahi2019, Li2022, Meng2023a} have shown that robust training may result in a significant decrease of the accuracy on benign samples, which is undesirable.

Few studies have explored the possibility of improving the machine learning performance using adversarial examples. For image classification, Xie \emph{et al.} \cite{Xie2020} employed a separate auxiliary batch normalization for adversarial examples to prevent model overfitting. For EEG classification, Ni \emph{et al.} \cite{Ni2022} used a loss on adversarial examples to improve the cross-subject and cross-state transfer learning performance. However, Li \emph{et al.} \cite{Li2022} and Meng \emph{et al.} \cite{Meng2023a} have shown that conventional robust training approaches usually lead to an evident reduction in BCI model accuracy on benign samples, i.e., it is difficult to achieve both high accuracy and good robustness through robust training.

Robust models aim to maintain good classification performance under adversarial attacks, which is important in safety-critical applications. However, the  accuracy degradation of robust models on benign samples seriously affects their adaption. To mitigate this problem, we propose alignment-based adversarial training (ABAT) to align EEG data for each session before performing robust training on them. This simple approach can be readily used in deep model training. After ABAT, the model's classification accuracy on benign samples and robustness on adversarial samples can simultaneously be improved. Experiments on five datasets using two different BCI paradigms, three classifiers and three different experimental settings demonstrated its effectiveness. To our knowledge, this is the first work on simultaneously improving the accuracy and robustness of the classifiers in EEG-based BCIs, and also the first time that EEG data alignment has been used in BCI adversarial defense. We hope that our findings can inspire more future research on robust EEG classifiers.

The remainder of this paper is organized as follows: Section~\ref{sect:rw} introduces related works. Section~\ref{sect:atad} proposes ABAT. Section~\ref{sect:es} describes the experimental settings. Section~\ref{sect:er} presents the experimental results. Finally, Section~\ref{sect:CFR} draws conclusions.

\section{Related Work} \label{sect:rw}

This section introduces background knowledge on EEG data alignment, adversarial attacks, and AT.

\subsection{Euclidean Alignment}

EEG data from different subjects/sessions can be regarded as data from different domains. Due to inter-subject/-session variations, the marginal probability distributions of EEG trials from different subjects/sessions are usually (significantly) different \cite{Zhang2020a}. Consequently, it is important to perform EEG data alignment to reduce the domain discrepancy.

Various EEG data alignment approaches have been proposed, which are reviewed and compared in \cite{Wu2022}. Zanini \emph{et al.} \cite{Zanini2018} introduced Riemannian alignment to align the covariance matrices of EEG trials from different subjects in the Riemannian space. He and Wu \cite{He2019} extended Riemannian alignment to Euclidean alignment (EA), which aligns the raw EEG trials in the Euclidean space. EA is efficient and completely unsupervised, demonstrating promising performance in different BCI paradigms \cite{drwuMITLBCI2022}.

For $N$ EEG trials $\{X_n\}_{n=1}^N$ in a particular domain, EA first computes the Euclidean arithmetic mean $\bar{R}$ of all $N$ spatial covariance matrices:
\begin{align}
\bar{R}=\frac{1}{N} \sum_{n=1}^{N} X_n\left(X_n\right)^{\top}.\label{alg:ea1}
\end{align}
Then, it performs the alignment by:
\begin{align}
\tilde{X}_n=\bar{R}^{-1 / 2} X_n, \quad n=1,...,N
\label{alg:ea2}
\end{align}

After EA, the aligned EEG trials $\{\tilde{X}_n\}_{n=1}^N$ in each domain are whitened, i.e., their average spatial covariance matrix becomes the identity matrix. Thus, EEG data distributions from different domains become more consistent.

\subsection{Incremental EA}

In online applications, target domain EEG trials $X^t$ arrive one by one on-the-fly, so there is a need to perform incremental EA on them.

Incremental EA applies EA to online EEG classification \cite{Li2024}. Let $\bar{R}_n^t$ be the average spatial covariance matrix computed from the first $n$ target domain EEG trials. When the $(n+1)$-th target domain EEG trial $X^t_{n+1}$ arrives, we first update:
\begin{align}
\bar{R}^t_{n+1}=\frac{1}{n+1} \left[n\bar{R}_n^t +X^t_{n+1}\left(X^t_{n+1}\right)^{\top}\right],\label{alg:ea3}
\end{align}
and then perform EA on $X^t_{n+1}$ using:
\begin{align}
\tilde{X}^t_{n+1}=(\bar{R}^t_{n+1})^{-1/2} X^t_{n+1}.\label{alg:ea4}
\end{align}

\subsection{Adversarial Attack}

Adversarial examples should closely resemble benign ones, achieved by imposing constraints on the perturbation magnitude. Let $D$ be a distance matric (a common choice is the $\ell_p$ norm), and $X^{adv}$ an adversarial example satisfying $D(X^{adv},X) < \epsilon$, where $\epsilon$ regulates the magnitude of the perturbation.

We consider the following three representative adversarial attack approaches:
\begin{enumerate}
\item Fast gradient sign method (FGSM) \cite{Goodfellow2015}, a straightforward yet highly effective adversarial attack strategy. It constructs an adversarial example through a single-step gradient computation:
	\begin{align}
	X^{adv}=X+{\epsilon}\cdot {\rm sign}\left(\nabla_{X}\mathcal{L}
	(\mathbf{\emph{C}_{\bm{\theta}}}(X),y)\right),\label{fgsm}
	\end{align}
	where $\emph{C}_{\bm{\theta}}$ is a classifier with parameter $\bm{\theta}$, and $\mathcal{L}$ its loss function. FGSM perturbs the input along the gradient direction, increasing the classifier's loss on its true label and leading to misclassification.

\item Projected Gradient Descent (PGD) \cite{Madry2018}, which is an iterative extension of FGSM. It starts from a perturbed version of the benign example $X$:
\begin{align}
    X_0^{adv} = X + \bm{\xi},\label{pgd1}
\end{align}
where $\bm{\xi}$ is uniform random noise sampled from $(-\epsilon,\epsilon)$. The iterative step is given by:
\begin{align}
    X_i^{adv}=&\text{Proj}_{X, \epsilon}\left(X_{i-1}^{adv}\right.\nonumber\\
     &\quad \left. + \alpha \cdot \text{sign}(\nabla_{X_{i-1}^{adv}} \mathcal{L}(\mathbf{C_{\bm{\theta}}}(X_{i-1}^{adv}), y))\right), \label{pgd2}
\end{align}
where $\alpha \leq \epsilon$ is the step size. The function $\text{Proj}_{X, \epsilon}$ ensures that $X_i^{adv}$ remains within the $\epsilon$-neighborhood of $X$ according to the $\ell_\infty$ norm.

\item AutoAttack \cite{Croce2020}, which combines four distinct attack strategies, including two budget-aware step size-free PGD variants with different losses (cross-entropy loss, and difference of logits ratio loss), square attack \cite{Andriushchenko2020}, and fast adaptive boundary attack \cite{Croce2020a}, each serving a unique purpose. AutoAttack is parameter-free, and has demonstrated superior performance in defeating various defense approaches \cite{Croce2020}.
\end{enumerate}

\subsection{AT}

AT \cite{Madry2018} is a classical robust training approach for enhancing the robustness of machine learning models against adversarial attacks, i.e., improving the model accuracy on adversarial examples by adding them to the training data. It can be expressed as a min-max (saddle point) optimization problem:
\begin{align}
\min_{\bm{\theta}} \mathbb{E}_{(X, y) \sim \mathcal{D}}\left[\max_{X^{adv} \in \mathcal{B}(X, \epsilon)} \mathcal{L}\left(C_{\bm{\theta}}\left(X^{adv}\right), y\right)\right],\label{algo:at}
\end{align}
where $\mathcal{D}$ is the data distribution, and $\mathcal{B}(X, \epsilon)$ the $\ell_\infty$ ball of radius $\epsilon$ centered at $X$. $X^{adv}$ can be adversarial examples generated by PGD \cite{Madry2018} (AT-PGD) or FGSM \cite{Wong2020} (AT-FGSM) or AutoAttack \cite{Croce2020}, etc.

Whereas AT can significantly enhance a model's robustness to adversarial examples, it often comes at the cost of reduced accuracy on benign examples \cite{Meng2023a}.

\section{Alignment-based Adversarial Training (ABAT)} \label{sect:atad}

AT aims at learning model parameters $\bm{\theta}$ that minimize the model's loss $\mathcal{L}$ on training samples' strong adversarial examples $X^{adv}$, i.e., to increase its robustness. AT is one of the most effective adversarial defense approaches. However, it often comes at the cost of reduced accuracies on benign examples. This paper studies whether AT can be used to improve the model's robustness and accuracy simultaneously.

Ni \emph{et al.} \cite{Ni2022} were the first to include the loss on adversarial examples in the overall model training loss function in EEG classification, to improve the transfer learning classification accuracy. However, they did not consider the adversarial robustness. Li \emph{et al.} \cite{Li2022} and Meng \emph{et al.} \cite{Meng2023a} showed that frequently, robust training, one of the most popular adversarial defense approaches, degrades the accuracy of BCI models. This may be due to the lack of EEG data alignment to reduce the data discrepancy among different subjects or different sessions. Multiple EEG data aliment approaches have been proposed, e.g., Riemannian alignment \cite{Zanini2018}, EA \cite{He2019} and label alignment \cite{drwuLA2020}. They greatly improve the classification accuracy in traditional transfer learning scenarios \cite{Wu2022}. However, EEG dat alignment has not been used in BCI adversarial defense.

\begin{figure}[htpb]\centering
{\includegraphics[width=\linewidth,clip]{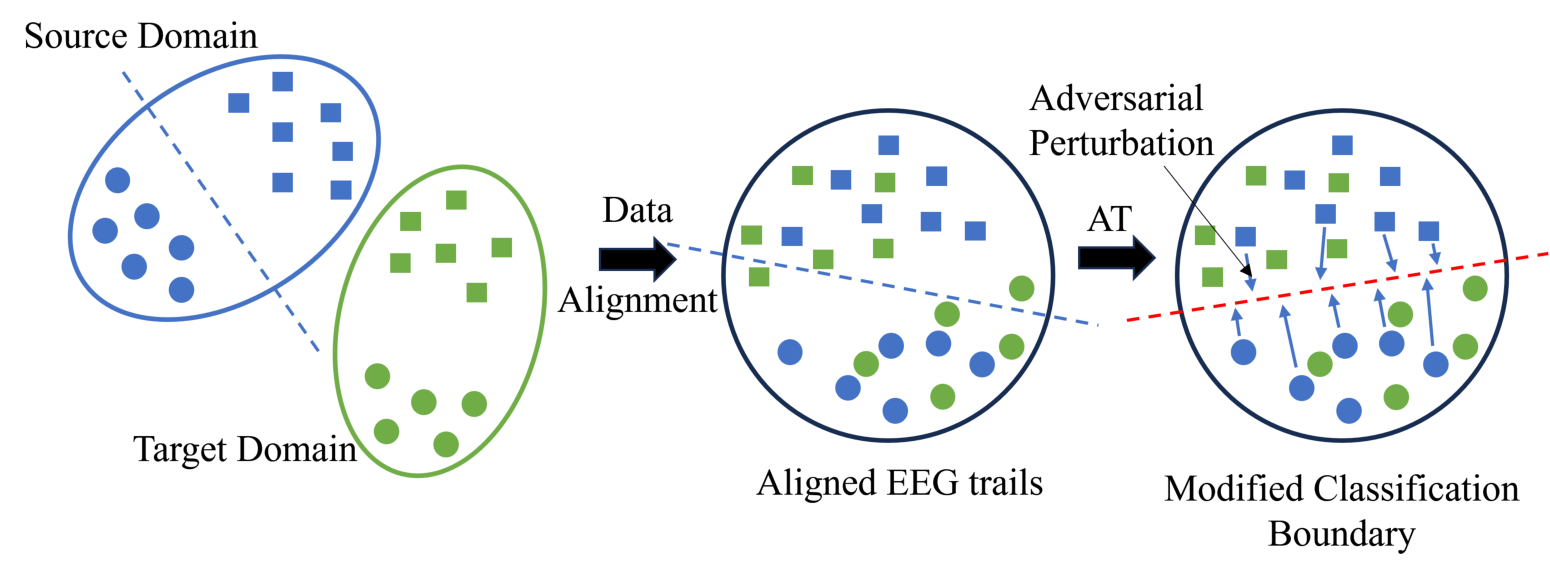}}
\caption{The influence of ABAT on EEG data from different domains. Data alignment aligns EEG trials from different domains to reduce their distribution discrepancies, and AT further robustifies the classification boundary.} \label{fig:eaat}
\end{figure}

We propose a very simple yet effective ABAT to fill this gap, by performing EEG data alignment before AT. Data alignment aligns EEG trials from different domains to reduce their distribution discrepancies, and AT further robustifies the classification boundary, as illustrated in Fig.~\ref{fig:eaat}. EA is used in this paper, for its simplicity and effectiveness.

Algorithm~\ref{Alg:ABAT} gives the pseudo-code of ABAT. It first aligns the EEG data of each domain using EA, and then performs adversarial training.

\begin{figure}[htpb]\centering
{\includegraphics[width=\linewidth,clip]{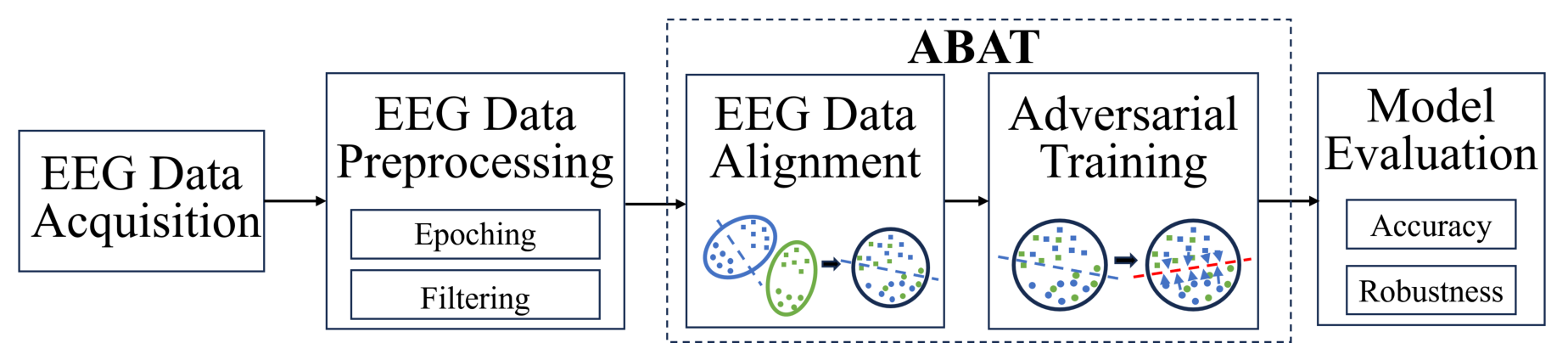}}
\caption{The complete BCI flowchart, incorporating ABAT. After preprocessing EEG data using epoching and filtering, ABAT trains the classifier, which is then used in subsequent classification and robustness evaluation. ABAT aligns EEG data centers across different domains and robustifies the classifier's decision boundary through adversarial training.} \label{fig:wp}
\end{figure}

The complete BCI flowchart with ABAT consists of data acquisition, data preprocessing (EEG data epoching and filtering), data alignment, AT, and model evaluation (in terms of accuracy and robustness), as shown in Fig.~\ref{fig:wp}. Particularly, after preprocessing the EEG data, ABAT is used to train the EEG classifier, which is then used in subsequent classification and robustness evaluation.

\begin{algorithm}[htpb]
	\caption{Alignment-based adversarial training (ABAT).}\label{Alg:ABAT}
	\KwIn{$\mathcal{S}=\{\mathcal{D}_s\}_{s=1}^{S}$, labeled data from $S$ source domains\;
		\hspace*{10mm} $M$, the number of model training epochs.}
	\KwOut{$C_{\bm{\theta}}$, trained EEG classifier.}
	
	Randomly initialize classifier $C_{\bm{\theta}}$ or pre-train the classifier $C_{\bm{\theta}}$ on other available data\;
	\tcp{Data Alignment}
	\For{$s=1:S$}{	
	Perform EA on $\mathcal{D}_s$ by (\ref{alg:ea1}) and (\ref{alg:ea2}) to obtain aligned EEG data $\tilde{\mathcal{D}}_s$;
 	}
 	Denote $\tilde{\mathcal{S}}=\{\tilde{\mathcal{D}}_s\}_{s=1}^{S}$\;
 	\tcp{AT}
 	\For{$m=1:M$}{	
 	Generate adversarial examples $\tilde{\mathcal{S}}^{adv}$ by (\ref{fgsm}) or by (\ref{pgd1}) and (\ref{pgd2}) or by AutoAttack on aligned source domain data $\tilde{\mathcal{S}}$ and $C_{\bm{\theta}}$\;
    Update the classifier $C_{\bm{\theta}}$ by (\ref{algo:at}) on generated adversarial examples $\tilde{\mathcal{S}}^{adv}$.
	}

\textbf{Return} $C_{\bm{\theta}}$.
\end{algorithm}

\section{Experiment Settings} \label{sect:es}

This section introduces our experiment settings, including the datasets, models, performance evaluation metrics, testing scenarios, and hyper-parameters. Our source code, including data preprocessing, can be found at https://github.com/xqchen914/ABAT.

\subsection{Datasets}

The following four datasets, summarized in Table~\ref{tab:datasets}, were used:

\begin{table}[htbp] \centering \setlength{\tabcolsep}{1mm}\centering
\caption{Summary of four datasets.}\label{tab:datasets}
\begin{tabular}{c|cccccc} \toprule
\multirow{2}{*}{Dataset}&\# of&\# of Time&\# of&Trails per& \# of &Class-  \\
&Subjects&Points&Channels&Subject&Sessions&Imbalance\\ \midrule
MI4&9&1000&22&576&2&\ding{55}\\
MI6&10&800&64&480&1&\ding{55}\\
MI2&60&512&27&240&1&\ding{55}\\
P300&8&128&32&3300   & 4&\checkmark\\
ERP&10&206&16&1728&3&\checkmark\\ \bottomrule
\end{tabular}
\end{table}

\begin{enumerate}
\item Four-class motor imagery dataset (MI4) \cite{Tangermann2012}: This is Dataset 2a in BCI Competition IV$\footnote{https://www.bbci.de/competition/iv/}$. It was collected from 9 subjects in two sessions on different days. There are four classes, i.e., left hand, right hand, feet, and tongue. The 22-channel EEG signals were sampled at 256 Hz. We extracted the data in [0, 4] seconds after each imagination prompt and band-pass filtered the trials at [8, 32] Hz. Each subject had 144 EEG epochs per class.
\item Six-class motor imagery dataset (MI6) \cite{Yi2014}: It was collected from 10 subjects for seven-class classification, i.e., left hand, right hand, feet, both hands, left hand combined with right foot, right hand combined with left foot, and rest state. The data collection process was divided into 9 sections, with 5 to 10 minutes intersection break. We only used data from the first six classes. The 64-channel EEG signals were sampled at 200 Hz. We extracted the data in [0, 4] seconds after each imagination prompt and band-pass filtered the trials at [4, 32] Hz. Each subject had 80 EEG epochs per class.
\item Two-class motor imagery (MI2) \cite{Dreyer2023} includes EEG data from 60 users performing left hand and right hand motor imagery tasks for 6 runs. 27-channel EEG signals were recorded at 512 Hz. After [4, 32] Hz band-pass filtering, we downsampled the data to 128 Hz and extracted data within [0, 4] seconds of each imagination prompt. Each subject had 120 EEG epochs per class.
\item P300 evoked potentials (P300) \cite{Hoffmann2008}: It was collected from four disabled subjects and four healthy ones in four sessions for two-class classification (target and non-target). The EEG data were recorded from 32 channels at 2048 Hz. We re-referenced the data, discarded the mastoid channels, filtered the data using a [1, 12] Hz bandpass filter, and down-sample the data to 128 Hz. EEG epochs between [0, 1] second were extracted. Each subject had 3,300 epochs, among which 557 were target.
\item Event related potential (ERP) \cite{Arico2014}: It was collected from 10 subjects in three sessions for two-class classification (target and non-target). The EEG signals were recorded using 16 electrodes at 250 Hz. We used MOABB API$\footnote{https://neurotechx.github.io/moabb/api.html}$ to get the preprocessed EEG data. EEG epochs between [0, 0.8] second were extracted. Each subject had 1,728 EEG epochs, among which about 288 were target.
\end{enumerate}

\subsection{Evaluation Metrics}

We used balanced classification accuracy (BCA) to evaluate the classification performance. The frequently-used raw classification accuracy (RCA) is the ratio of the number of correctly classified examples to the number of total examples. The BCA is the average of the per-class RCAs. BCA was preferred in our experiments because the ERP and P300 datasets have significant intrinsic class imbalance, so using RCA is misleading. When all classes have the same number of samples (e.g., MI4 and MI6), BCA reduces to RCA.

We repeated each experiment three times, and their averages were reported.

\subsection{Evaluated Classifiers}

The following three CNN classifiers were used:
\begin{enumerate}
\item EEGNet \cite{Lawhern2018}: It is a compact CNN architecture specifically designed for EEG classification tasks. This model comprises two convolutional blocks and a single classification block. It employs depthwise and separable convolutions, as opposed to conventional convolutions, to effectively reduce the model's parameter size.
\item DeepCNN \cite{Schirrmeister2017}: In contrast to EEGNet, DeepCNN has a higher number of parameters. It consists of three convolutional blocks and a softmax layer for classification. The first convolutional block is customized for EEG inputs, and the last two are standard ones.
\item ShallowCNN \cite{Schirrmeister2017}: ShallowCNN is a simplified variant of DeepCNN, inspired from filter bank common spatial patterns. Compared with DeepCNN, ShallowCNN has a convolutional block with a larger kernel, a different activation function, and different pooling techniques.
\end{enumerate}

If not specified otherwise, the three convolutional blocks of DeepCNN used 25, 50 and 100 convolutional kernels, respectively, and the convolutional block of ShallowCNN used 40 convolutional kernels. EEGNet, DeepCNN and ShallowCNN had 1676, 94,079 and 57,804 parameters, respectively. In terms of model capacity, EEGNet<ShallowCNN<DeepCNN.

\subsection{Testing Scenarios}

We tested the performance of different models under adversarial attacks, i.e., their robustness, in offline scenario, where adversarial attacks are most effective. We also tested their classification accuracies on benign samples in both offline and online scenarios. Algorithms~\ref{Alg:testABAT} gives the pseudo-code of the corresponding online test, offline test, and offline robust test procedures.

\begin{algorithm}[htpb]
	\caption{Online test, offline test, and offline robustness test procedures.}\label{Alg:testABAT}
	\KwIn{$T$ target sessions/domains $\{X^{t}_n\}_{n=1}^{N_t}$, $t = 1,...,T$\;
		\hspace*{10mm} $f$, the classifier.}
	\KwOut{$\hat{y}^{t}_n$, classification results for $X^{t}_n$\;
			\hspace*{11mm} $\hat{y}^{t,adv}_n$, classification results for adversarial
			\hspace*{11mm} counterparts of $X^{t}_n$.}
	\tcp{Online test}
	
	\For{$t=1:T$}{	
		\For{$n=1:N_t$}{	
		Perform incremental EA on $X^{t}_n$ by (\ref{alg:ea3}) and (\ref{alg:ea4}) to obtain $\tilde{X}^{t}_n$\;
		Compute $\hat{y}^{t}_n=f(\tilde{X}^{t}_n)$\;
		}
	}
	
	\tcp{Offline test}
	
	\For{$t=1:T$}{	
	Perform EA on $\{X^{t}_n\}_{n=1}^{N_t}$ by (\ref{alg:ea1}) and (\ref{alg:ea2}) to obtain $\{\tilde{X}^{t}_n\}_{n=1}^{N_t}$\;
	}
	Compute $\hat{y}^{t}_n=f(\tilde{X}^{t}_n)$\;
	
	\tcp{Offline robustness test}
	
	Compute adversarial examples $\{\hat{X}^{t,adv}_n\}_{n=1}^{N_t}$ of $\{\hat{X}^{t}_n\}_{n=1}^{N_t}$ by (\ref{fgsm}), or (\ref{pgd1}) and (\ref{pgd2}), or AutoAttack \cite{Croce2020}\;
	Compute $\hat{y}^{t,adv}_n=f(\tilde{X}^{t,adv}_n)$.
\end{algorithm}

On MI4 and ERP, we used the first session as the training set and the remaining ones as the test set. On P300, we used the first two sessions as the training set. On MI6 and MI2, we used the first two blocks of data as the training set.

\subsection{Hyper-Parameters}

When training within-subject models, batch size 32 was used on MI4, MI6 and ERP datasets, and batch size 128 on the P300 dataset. When training cross-subject models, batch size 128 was used on all four datasets. All models were trained for 100 epochs with initial learning rate 0.01, which was reduced to 0.001 after 50 epochs.

Since the perturbation magnitude is correlated with the original EEG signal magnitude, we selected the perturbation magnitude to be $\epsilon$ times of the EEG signal standard deviation. In adversarial attacks, we chose 20 iterations for PGD and AutoAttack, with attack step size $\epsilon/10$ for PGD. In AT and ABAT, we used 10 iterations for PGD, with PGD attack step size $\alpha=\epsilon/5$.

Deep learning models perform very differently on adversarial examples with different perturbation amplitudes \cite{Meng2023a}. To have a more comprehensive assessment, we evaluated the model performance with different adversarial training perturbation amplitudes on adversarial examples with different perturbation amplitudes.

We implemented AT-FGSM, AT-PGD, ABAT-FGSM and ABAT-PGD with various training perturbation magnitudes $\epsilon$ on the four datasets. More specifically, $\epsilon=0.01$ was used in AT-FGSM and AT-PGD on MI4, MI6 and P300, and $\epsilon\in\{0.01,0.03\}$ on ERP. $\epsilon\in\{0.01,0.03,0.05\}$ were used in ABAT-FGSM and ABAT-PGD on MI4, MI6 and P300, and $\epsilon\in\{0.01, 0.03, 0.05, 0.07, 0.09\}$ on ERP. We calculated the resulting classifiers' BCAs for benign samples, and under FGSM, PGD and AutoAttack adversarial attacks with $\epsilon=\{0.01, 0.03, 0.05\}$.

\section{Experimental Results} \label{sect:er}

This section presents experimental results to verify the effectiveness of our proposed ABAT.

\subsection{Offline Cross-Block/-Session Performance on Benign Samples} \label{sect:ma}

Offline cross-block/-session BCAs of EEGNet, DeepCNN and ShallowCNN under benign training (BT), AT-FGSM and AT-PGD, and ABAT-FGSM and ABAT-PGD with different training perturbation magnitudes $\epsilon$, on the benign samples of the five datasets are shown in the `No Attack' column of Tables~\ref{tab:MI4_within}-\ref{tab:ERP_within}, respectively. Observe that:
\begin{enumerate}
	\item Without EA, AT-FGSM and AT-PGD had similar or lower BCAs on the benign samples than BT.
	\item With EA, as ABAT-FGSM or ABAT-PGD perturbation amplitude increased from 0.01 to 0.05, the BCAs on the benign samples first increased and then decreased, or kept increasing.
	\item On MI4, MI6, MI2 and P300, EA greatly improved the BCAs of BT, and ABAT-FGSM and ABAT-PGD further improved the BCAs, regardless of the perturbation magnitude. On the ERP dataset, although EA decreased the BCAs, ABAT-FGSM and ABAT-PGD still outperformed BT.
	\item On the MI datasets, networks with larger capacities achieved their best performance on benign samples under larger ABAT perturbation amplitudes. More specifically, on MI4, EEGNet achieved its highest BCA on benign samples at $\epsilon=0.01$, whereas ShallowCNN and DeepCNN at $\epsilon=0.03$; on MI6, EEGNet reached its highest BCA on benign samples at $\epsilon=0.01$ (PGD) or $\epsilon=0.03$ (FGSM), whereas ShallowCNN at $\epsilon=0.03$ (PGD) or $\epsilon=0.05$ (FGSM), and DeepCNN at $\epsilon=0.05$; on MI2, EEGNet achieved its highest BCA on benign samples at $\epsilon=0.03$, whereas ShallowCNN and DeepCNN at $\epsilon=0.05$. This difference was not obvious on the P300 and ERP datasets, probably due to the intrinsic differences of BCI paradigms.
\end{enumerate}

\subsection{Offline Cross-Block/-Session Performance on Adversarial Examples}

Offline cross-block/-session BCAs of EEGNet, DeepCNN and ShallowCNN under BT, AT-FGSM and AT-PGD, and ABAT-FGSM and ABAT-PGD with different training perturbation magnitudes $\epsilon$, under FGSM, PGD and AutoAttack adversarial attacks on the five datasets are shown in Tables~\ref{tab:MI4_within}-\ref{tab:ERP_within}. Observe that:
\begin{enumerate}
\item AT-FGSM, AT-PGD, ABAT-FGSM and ABAT-PGD with different perturbation amplitudes greatly improved the BCAs on the adversarial examples, showing good generalization.
\item When the perturbation amplitude of BAT-FGSM and ABAT-PGD increased, the BCAs for adversarial examples with larger perturbation amplitudes also increased, but the BCAs for adversarial examples with small perturbation amplitude may be reduced. Many times there was a trade-off between the BCAs on adversarial examples with large perturbation amplitude and those with small perturbation amplitude.
\item The overall performance of ABAT-FGSM and ABAT-PGD on the three classifiers and four datasets was similar. However, ABAT-FGSM is much faster than ABAT-PGD.
\end{enumerate}

\begin{table*}[htbp] \centering \setlength{\tabcolsep}{2mm} \scriptsize
\caption{BCAs of different training approaches under benign samples and various attacks on MI4.} \label{tab:MI4_within}
\begin{tabular}{c|c|cccccccccccc} \toprule
\multirow{2}{*}{Model}
&\multirow{2}{*}{EA}&\multirow{2}{*}{Training}
&No&FGSM &FGSM &FGSM &PGD &PGD &PGD& AutoAttack& AutoAttack& AutoAttack&\multirow{2}{*}{\makebox[0.03\textwidth][c]{Avg.}}\\
&&&Attack&0.01&0.03&0.05&0.01&0.03&0.05&0.01&0.03&0.05 \\ \midrule
\multirow{10}{*}{EEGNet}
 & \multirow{3}{*}{\shortstack[c]{w/o \\ EA}}&BT                 & 60.78 & 36.14 & 8.89  & 1.66  & 35.92 & 7.70  & 1.05  & 35.64 & 7.32  & 0.94  & 19.60 \\
 & &AT-FGSM 0.01 & 56.71 & 48.95 & 34.10 & 22.61 & 49.02 & 33.74 & 21.63 & 48.87 & 33.51 & 21.17 & 37.03 \\
 & &AT-PGD 0.01      & 55.99 & 48.33 & 34.25 & 22.31 & 48.47 & 34.03 & 21.37 & 48.28 & 33.71 & 20.95 & 36.77 \\ \cmidrule(r){2-14}
& \multirow{7}{*}{\shortstack[c]{with \\ EA }} &BT          & 69.79          & 57.54          & 35.11          & 19.98          & 57.50          & 34.74          & 19.01          & 57.46           & 34.49           & 18.45           & 40.41        \\
           && ABAT-FGSM 0.01 & 71.08          & 64.11          & 49.52          & 35.65          & 64.12          & 49.28          & 35.11          & 64.08           & 49.10           & 34.70           & 51.67          \\
           && ABAT-FGSM 0.03 & 68.92          & 65.33          & 56.71          & 48.25          & 65.33          & 56.61          & 48.11          & 65.33           & 56.58           & 47.90           & 57.91          \\
           && ABAT-FGSM 0.05 & 65.46          & 62.91          & 57.29          & 51.34          & 62.92          & 57.23          & 51.35          & 62.91           & 57.20           & 51.16           & 57.98          \\
           && ABAT-PGD 0.01  & \textbf{73.28} & 66.82          & 51.53          & 37.49          & 66.80          & 51.29          & 36.70          & \textbf{66.78}  & 51.09           & 36.34           & 53.81          \\
           && ABAT-PGD 0.03  & 70.69          & \textbf{66.85} & 58.81          & 49.76          & \textbf{66.85} & 58.83          & 49.50          & 66.85           & 58.74           & 49.31           & 59.62          \\
           && ABAT-PGD 0.05  & 67.48          & 64.75          & \textbf{59.26} & \textbf{53.16} & 64.75          & \textbf{59.25} & \textbf{53.15} & 64.74           & \textbf{59.19}  & \textbf{53.00}  & \textbf{59.87} \\  \midrule

\multirow{10}{*}{DeepCNN}
 & \multirow{3}{*}{\shortstack[c]{w/o \\ EA}}& BT    & 54.06 & 31.61 & 6.93  & 1.02  & 31.29 & 6.30  & 0.82  & 31.08 & 5.97  & 0.68  & 16.98 \\
 & &AT-FGSM 0.01& 49.19 & 41.82 & 27.87 & 16.23 & 41.74 & 27.47 & 15.65 & 41.69 & 27.22 & 15.30 & 30.42 \\
 & &AT-PGD 0.01& 49.06 & 41.53 & 27.69 & 16.54 & 41.51 & 27.40 & 15.95 & 41.49 & 27.26 & 15.63 & 30.41 \\ \cmidrule(r){2-14}
& \multirow{7}{*}{\shortstack[c]{with \\ EA}}& BT  & 59.70          & 44.21          & 20.41          & 7.70           & 43.98          & 19.43          & 6.42           & 43.84           & 18.88           & 5.93            & 27.05          \\
           && ABAT-FGSM 0.01 & 62.73          & 53.20          & 35.08          & 21.58          & 53.15          & 34.57          & 20.78          & 53.15           & 34.32           & 20.37           & 38.89          \\
           && ABAT-FGSM 0.03 & \textbf{66.05} & 58.82          & 44.71          & 32.33          & 58.82          & 44.55          & 31.71          & 58.80           & 44.37           & 31.44           & 47.16          \\
          & & ABAT-FGSM 0.05 & 63.88          & \textbf{59.08} & \textbf{48.74} & \textbf{38.71} & \textbf{59.08} & \textbf{48.66} & \textbf{38.49} & \textbf{59.08}  & \textbf{48.60}  & \textbf{38.28}  & \textbf{50.26} \\
           && ABAT-PGD 0.01  & 61.54          & 52.28          & 34.23          & 21.14          & 52.20          & 33.77          & 20.22          & 52.17           & 33.56           & 19.88           & 38.10          \\
           && ABAT-PGD 0.03  & 64.98          & 58.64          & 44.82          & 32.72          & 58.60          & 44.68          & 32.19          & 58.59           & 44.60           & 31.88           & 47.17          \\
           && ABAT-PGD 0.05  & 62.83          & 58.00          & 48.01          & 38.25          & 58.00          & 47.92          & 37.96          & 57.97           & 47.81           & 37.80           & 49.45          \\ \midrule

\multirow{10}{*}{ShallowCNN}
 & \multirow{3}{*}{\shortstack[c]{w/o \\ EA}} &BT   & 60.57 & 31.92 & 7.27  & 1.08  & 31.60 & 6.11  & 0.73  & 31.48 & 5.80  & 0.57  & 17.71 \\
 & &AT-FGSM 0.01& 57.47 & 47.15 & 29.57 & 16.87 & 47.13 & 29.37 & 16.00 & 47.08 & 29.05 & 15.46 & 33.51 \\
 & &AT-PGD 0.01& 57.47 & 47.49 & 29.73 & 16.80 & 47.48 & 29.64 & 16.06 & 47.44 & 29.24 & 15.44 & 33.68 \\ \cmidrule(r){2-14}
& \multirow{7}{*}{\shortstack[c]{with \\ EA}}&BT & 71.46          & 57.01          & 34.40          & 19.24          & 56.94          & 33.90          & 18.24          & 56.91           & 33.74           & 17.75           & 39.96          \\
           && ABAT-FGSM 0.01 & 73.82          & 65.41          & 48.83          & 34.45          & 65.35          & 48.68          & 33.90          & 65.34           & 48.61           & 33.71           & 51.81          \\
           && ABAT-FGSM 0.03 & 74.79          & 68.58          & 54.99          & 42.48          & 68.57          & 54.85          & 42.21          & 68.56           & 54.81           & 42.08           & 57.19          \\
           && ABAT-FGSM 0.05 & 73.88          & 68.36          & \textbf{57.42} & \textbf{45.88} & 68.35          & \textbf{57.39} & \textbf{45.82} & 68.35           & \textbf{57.38}  & \textbf{45.64}  & \textbf{58.85} \\
           && ABAT-PGD 0.01  & 73.95          & 65.61          & 48.70          & 34.30          & 65.59          & 48.47          & 33.69          & 65.59           & 48.34           & 33.38           & 51.76          \\
           && ABAT-PGD 0.03  & \textbf{74.99} & \textbf{68.85} & 55.65          & 42.26          & \textbf{68.85} & 55.56          & 41.96          & \textbf{68.84}  & 55.52           & 41.67           & 57.41          \\
           && ABAT-PGD 0.05  & 73.51          & 68.17          & 56.97          & 45.74          & 68.16          & 56.94          & 45.63          & 68.16           & 56.93           & 45.46           & 58.57          \\
\bottomrule
\end{tabular}
\end{table*}

\begin{table*}[htbp] \centering \setlength{\tabcolsep}{2mm}
\scriptsize
\caption{BCAs of different training approaches under benign samples and various attacks on MI6.}\label{tab:MI6_within}
\begin{tabular}{c|c|cccccccccccc} \toprule
\multirow{2}{*}{Model} &\multirow{2}{*}{EA}
&\multirow{2}{*}{Training}
&No&FGSM &FGSM &FGSM &PGD &PGD &PGD& AutoAttack& AutoAttack& AutoAttack&\multirow{2}{*}{\makebox[0.03\textwidth][c]{Avg.}}\\
&&&Attack&0.01&0.03&0.05&0.01&0.03&0.05&0.01&0.03&0.05 \\ \midrule
\multirow{10}{*}{EEGNet}
 & \multirow{3}{*}{\shortstack[c]{w/o \\ EA}}&BT    & 29.52 & 16.75 & 4.91  & 1.80  & 16.71 & 4.71  & 1.52  & 16.61 & 4.44  & 1.42  & 9.84  \\
 & &AT-FGSM 0.01 & 29.69 & 25.60 & 18.50 & 12.77 & 25.59 & 18.49 & 12.59 & 25.57 & 18.40 & 12.30 & 19.95 \\
 & &AT-PGD 0.01& 28.38 & 24.79 & 17.50 & 12.24 & 24.77 & 17.45 & 12.08 & 24.76 & 17.39 & 11.99 & 19.14 \\  \cmidrule(r){2-14}
 & \multirow{7}{*}{\shortstack[c]{with \\ EA}}&BT & 32.71          & 9.67           & 0.80           & 0.06           & 9.40           & 0.61           & 0.04           & 9.21            & 0.52            & 0.03            & 6.31           \\
           && ABAT-FGSM 0.01 & 41.13          & 29.81          & 13.69          & 6.26           & 29.78          & 13.40          & 5.78           & 29.73           & 13.18           & 5.52            & 18.83          \\
           && ABAT-FGSM 0.03 & \textbf{42.91} & \textbf{38.24} & 30.37          & 22.93          & \textbf{38.23} & 30.30          & 22.94          & \textbf{38.23}  & 30.25           & 22.57           & 31.70          \\
           && ABAT-FGSM 0.05 & 40.57          & 37.86          & \textbf{32.89} & \textbf{28.01} & 37.85          & \textbf{32.97} & \textbf{28.02} & 37.85           & \textbf{32.86}  & \textbf{27.91}  & \textbf{33.68} \\
           && ABAT-PGD 0.01  & 40.98          & 29.54          & 12.69          & 5.09           & 29.54          & 12.45          & 4.67           & 29.44           & 12.16           & 4.30            & 18.09          \\
           && ABAT-PGD 0.03  & 40.28          & 36.07          & 28.18          & 21.24          & 36.07          & 28.15          & 21.16          & 36.07           & 28.00           & 20.97           & 29.62          \\
           && ABAT-PGD 0.05  & 38.29          & 35.84          & 30.72          & 26.70          & 35.84          & 30.74          & 26.67          & 35.84           & 30.65           & 26.56           & 31.79          \\ \midrule

\multirow{10}{*}{DeepCNN}&
 \multirow{3}{*}{\shortstack[c]{w/o \\ EA}}& BT   & 21.96 & 15.78 & 7.46  & 3.04 & 15.73 & 7.28  & 2.83 & 15.72 & 7.23  & 2.71 & 9.97  \\
 & &AT-FGSM 0.01& 22.22 & 17.63 & 10.09 & 5.57 & 17.60 & 9.96  & 5.41 & 17.59 & 9.93  & 5.27 & 12.13 \\
 & &AT-PGD 0.01& 22.21 & 18.01 & 10.38 & 5.40 & 18.01 & 10.27 & 5.25 & 18.01 & 10.23 & 5.11 & 12.29 \\ \cmidrule(r){2-14}
& \multirow{7}{*}{\shortstack[c]{with \\ EA}}& BT          & 25.08          & 9.69           & 0.81           & 0.05           & 9.52           & 0.66           & 0.03           & 9.45            & 0.56            & 0.01            & 5.59           \\
           && ABAT-FGSM 0.01 & 27.76          & 14.18          & 2.44           & 0.33           & 14.08          & 2.27           & 0.27           & 14.01           & 2.08            & 0.21            & 7.76           \\
           && ABAT-FGSM 0.03 & 31.21          & 20.78          & 7.94           & 2.64           & 20.75          & 7.67           & 2.36           & 20.71           & 7.57            & 2.21            & 12.38          \\
           && ABAT-FGSM 0.05 & \textbf{34.54} & 26.43          & \textbf{14.88} & \textbf{8.14}  & 26.41          & \textbf{14.68} & \textbf{7.92}  & 26.39           & \textbf{14.64}  & \textbf{7.69}   & \textbf{18.17} \\
           && ABAT-PGD 0.01  & 27.51          & 14.52          & 2.66           & 0.25           & 14.45          & 2.42           & 0.19           & 14.41           & 2.24            & 0.18            & 7.88           \\
           && ABAT-PGD 0.03  & 32.07          & 21.49          & 8.12           & 2.87           & 21.47          & 7.86           & 2.63           & 21.47           & 7.78            & 2.51            & 12.83          \\
           && ABAT-PGD 0.05  & 34.17          & \textbf{26.57} & 14.81          & 7.93           & \textbf{26.55} & 14.61          & 7.68           & \textbf{26.54}  & 14.54           & 7.58            & 18.10          \\\midrule

\multirow{10}{*}{ShallowCNN}
 & \multirow{3}{*}{\shortstack[c]{w/o \\ EA}}& BT    & 35.69 & 22.10 & 7.31  & 2.88  & 22.08 & 7.18  & 2.75  & 22.03 & 7.07  & 2.63  & 13.17 \\
 & &AT-FGSM 0.01& 34.62 & 28.32 & 17.90 & 10.19 & 28.31 & 17.80 & 9.97  & 28.31 & 17.69 & 9.81  & 20.29 \\
 & & AT-PGD 0.01& 34.80 & 28.41 & 17.94 & 10.49 & 28.40 & 17.86 & 10.31 & 28.40 & 17.79 & 10.14 & 20.45 \\ \cmidrule(r){2-14}
& \multirow{7}{*}{\shortstack[c]{with \\ EA}} &BT           & 46.45          & 21.44          & 4.11           & 0.70           & 21.31          & 3.83           & 0.59           & 21.19           & 3.67            & 0.48            & 12.38          \\
           && ABAT-FGSM 0.01 & 49.77          & 36.72          & 17.92          & 8.26           & 36.66          & 17.71          & 7.79           & 36.66           & 17.60           & 7.62            & 23.67          \\
           && ABAT-FGSM 0.03 & 50.65          & 42.14          & 27.41          & 16.95          & 42.11          & 27.34          & 16.67          & 42.11           & 27.32           & 16.56           & 30.92          \\
           && ABAT-FGSM 0.05 & 50.72          & 43.68          & 31.53          & 22.03          & 43.69          & 31.49          & 21.84          & 43.68           & 31.45           & 21.73           & 34.18          \\
           && ABAT-PGD 0.01  & 49.92          & 36.90          & 18.23          & 8.67           & 36.89          & 17.97          & 8.36           & 36.88           & 17.89           & 8.12            & 23.98          \\
           && ABAT-PGD 0.03  & \textbf{51.44} & 42.09          & 27.71          & 17.19          & 42.10          & 27.61          & 16.92          & 42.08           & 27.56           & 16.78           & 31.15          \\
           && ABAT-PGD 0.05  & 51.16          & \textbf{44.11} & \textbf{31.79} & \textbf{22.04} & \textbf{44.09} & \textbf{31.74} & \textbf{21.86} & \textbf{44.09}  & \textbf{31.71}  & \textbf{21.74}  & \textbf{34.43}\\
\bottomrule
\end{tabular}
\end{table*}

\begin{table*}[htbp]
  \centering
  \caption{BCAs of different training approaches under benign samples and various attacks on MI2.}
    \begin{tabular}{c|c|ccccccccccc}
    \toprule
    \multirow{2}{*}{Model} & \multirow{2}{*}{EA}&\multirow{2}{*}{Training} & No  & FGSM & FGSM & FGSM & PGD & PGD & PGD & AutoAttack & AutoAttack & AutoAttack \\
      &   && Attack & 0.01 & 0.03 & 0.05 & 0.01 & 0.03 & 0.05 & 0.01 & 0.03 & 0.05 \\
    \midrule
    \multirow{10}[4]{*}{EEGNet} & \multirow{3}{*}{\shortstack[c]{w/o \\ EA}}& BT  & 64.22 & 43.05 & 17.32 & 7.49 & 43.02 & 16.99 & 7.00 & 43.01 & 16.93 & 6.92 \\
      & &AT-FGSM 0.01 & 64.34 & 57.44 & 43.16 & 30.29 & 57.44 & 43.07 & 29.89 & 57.44 & 43.04 & 29.82 \\
      & &AT-PGD 0.01 & 63.94 & 57.28 & 43.43 & 30.76 & 57.28 & 43.38 & 30.43 & 57.28 & 43.36 & 30.37 \\
\cmidrule{2-13}
& \multirow{7}{*}{\shortstack[c]{with \\ EA}}& BT  & 64.96 & 47.68 & 23.28 & 12.12 & 47.66 & 23.10 & 11.77 & 47.66 & 23.06 & 11.72 \\
      & &ABAT-FGSM 0.01 & 67.53 & 56.87 & 37.46 & 23.72 & 56.86 & 37.31 & 23.28 & 56.86 & 37.29 & 23.21 \\
      & &ABAT-FGSM 0.03 & \textbf{71.02} & 65.12 & 53.33 & 42.73 & 65.11 & 53.28 & 42.57 & 65.11 & 53.28 & 42.55 \\
      & &ABAT-FGSM 0.05 & 70.98 & \textbf{67.18} & \textbf{59.47} & \textbf{52.06} & \textbf{67.18} & \textbf{59.44} & \textbf{51.94} & \textbf{67.18} & \textbf{59.44} & \textbf{51.92} \\
      & &ABAT-PGD 0.01 & 68.36 & 57.42 & 37.56 & 23.47 & 57.41 & 37.40 & 23.13 & 57.41 & 37.38 & 23.07 \\
      & &ABAT-PGD 0.03 & 70.81 & 65.05 & 53.53 & 42.49 & 65.05 & 53.48 & 42.32 & 65.05 & 53.48 & 42.30 \\
      & &ABAT-PGD 0.05 & 70.06 & 66.62 & 59.17 & 51.51 & 66.62 & 59.16 & 51.38 & 66.62 & 59.16 & 51.37 \\
    \midrule
    \multirow{10}[4]{*}{DeepCNN} & \multirow{3}{*}{\shortstack[c]{w/o \\ EA}} & BT  & 57.31 & 48.90 & 33.05 & 19.48 & 48.89 & 32.95 & 19.32 & 48.88 & 32.91 & 19.30 \\
      & &AT-FGSM 0.01 & 57.95 & 51.89 & 39.64 & 28.63 & 51.89 & 39.61 & 28.51 & 51.89 & 39.59 & 28.48 \\
      & &AT-PGD 0.01 & 57.47 & 51.33 & 39.25 & 27.76 & 51.33 & 39.22 & 27.62 & 51.33 & 39.21 & 27.59 \\
\cmidrule{2-13}
& \multirow{7}{*}{\shortstack[c]{with \\ EA}}& BT  & 68.37 & 56.88 & 35.90 & 21.48 & 56.87 & 35.79 & 21.24 & 56.87 & 35.75 & 21.18 \\
      & &ABAT-FGSM 0.01 & 69.74 & 60.96 & 43.41 & 29.67 & 60.95 & 43.30 & 29.40 & 60.94 & 43.24 & 29.35 \\
      & &ABAT-FGSM 0.03 & 73.85 & 67.48 & 54.37 & 42.37 & 67.48 & 54.27 & 42.12 & 67.48 & 54.26 & 42.03 \\
      & &ABAT-FGSM 0.05 & 77.07 & \textbf{72.66} & \textbf{62.24} & 51.88 & \textbf{72.65} & \textbf{62.16} & 51.63 & \textbf{72.65} & \textbf{62.13} & 51.57 \\
      & &ABAT-PGD 0.01 & 70.30 & 61.74 & 44.56 & 30.17 & 61.72 & 44.45 & 29.90 & 61.72 & 44.41 & 29.82 \\
      & &ABAT-PGD 0.03 & 74.53 & 68.45 & 55.10 & 42.67 & 68.43 & 55.02 & 42.37 & 68.43 & 54.98 & 42.30 \\
      & &ABAT-PGD 0.05 & \textbf{77.15} & 72.36 & 62.19 & \textbf{52.22} & 72.36 & 62.10 & \textbf{51.95} & 72.36 & 62.09 & \textbf{51.88} \\
    \midrule
    \multirow{10}[4]{*}{ShallowCNN} &\multirow{3}{*}{\shortstack[c]{w/o \\ EA}} & BT  & 73.58 & 44.67 & 13.87 & 4.93 & 44.39 & 13.08 & 4.33 & 44.32 & 12.95 & 4.24 \\
      && AT-FGSM 0.01 & 73.75 & 64.67 & 43.96 & 25.86 & 64.65 & 43.64 & 25.10 & 64.65 & 43.45 & 24.71 \\
      && AT-PGD 0.01 & 73.60 & 64.12 & 43.61 & 25.70 & 64.11 & 43.36 & 24.97 & 64.11 & 43.23 & 24.60 \\
\cmidrule{2-13}      &\multirow{7}{*}{\shortstack[c]{with \\ EA}}& BT  & 74.26 & 39.42 & 12.86 & 6.32 & 39.28 & 12.50 & 5.89 & 39.28 & 12.43 & 5.82 \\
      & &ABAT-FGSM 0.01 & 76.42 & 64.93 & 42.15 & 26.03 & 64.92 & 41.90 & 25.21 & 64.91 & 41.68 & 24.53 \\
      & &ABAT-FGSM 0.03 & 77.94 & 69.35 & 51.12 & 36.14 & 69.35 & 51.03 & 35.90 & 69.35 & 51.02 & 35.78 \\
      & &ABAT-FGSM 0.05 & \textbf{77.98} & 70.71 & 54.87 & 41.00 & 70.71 & 54.85 & 40.86 & 70.71 & 54.84 & 40.84 \\
      & &ABAT-PGD 0.01 & 76.46 & 65.14 & 41.95 & 25.55 & 65.12 & 41.63 & 24.74 & 65.11 & 41.42 & 24.00 \\
      & &ABAT-PGD 0.03 & 77.80 & 69.10 & 50.84 & 35.98 & 69.10 & 50.78 & 35.75 & 69.10 & 50.77 & 35.65 \\
      & &ABAT-PGD 0.05 & 78.09 & \textbf{70.74} & \textbf{54.96} & \textbf{41.24} & \textbf{70.74} & \textbf{54.94} & \textbf{41.11} & \textbf{70.74} & \textbf{54.94} & \textbf{41.09} \\
    \bottomrule
    \end{tabular}%
  \label{tab:addlabel}%
\end{table*}%

\begin{table*}[htbp] \centering \setlength{\tabcolsep}{2mm}\scriptsize
\caption{BCAs of different training approaches under benign samples and various attacks on P300.}\label{tab:P300_within}
\begin{tabular}{c|c|cccccccccccc} \toprule
\multirow{2}{*}{Model}&\multirow{2}{*}{EA}
&\multirow{2}{*}{Training}
&No&FGSM &FGSM &FGSM &PGD &PGD &PGD& AutoAttack& AutoAttack& AutoAttack&\multirow{2}{*}{\makebox[0.03\textwidth][c]{Avg.}}\\
&&&Attack&0.01&0.03&0.05&0.01&0.03&0.05&0.01&0.03&0.05 \\ \midrule
\multirow{10}{*}{EEGNet}
 & \multirow{3}{*}{\shortstack[c]{w/o \\ EA}}& BT   & 61.39 & 58.29 & 50.83 & 44.00 & 58.29 & 50.80 & 43.95 & 58.29 & 50.80 & 43.94 & 52.06 \\
 & &AT-FGSM 0.01& 60.74 & 58.52 & 54.19 & 49.85 & 58.51 & 54.18 & 49.80 & 58.51 & 54.18 & 49.79 & 54.83 \\
 & &AT-PGD 0.01& 60.37 & 58.37 & 54.21 & 49.88 & 58.36 & 54.19 & 49.85 & 58.35 & 54.19 & 49.84 & 54.76 \\ \cmidrule(r){2-14}
& \multirow{7}{*}{\shortstack[c]{with \\ EA}}& BT         & 67.23          & 61.50          & 50.53          & 38.98          & 61.49          & 50.46          & 38.67          & 61.49           & 50.43           & 38.66           & 51.95          \\
           && ABAT-FGSM 0.01 & 69.60          & 65.60          & 57.36          & 48.81          & 65.60          & 57.33          & 48.61          & 65.60           & 57.32           & 48.61           & 58.44          \\
           && ABAT-FGSM 0.03 & \textbf{69.94} & 67.30          & 62.44          & 57.11          & 67.30          & 62.43          & 57.06          & 67.30           & 62.43           & 57.05           & 63.04          \\
           && ABAT-FGSM 0.05 & 68.98          & 67.24          & \textbf{63.61} & 59.71          & 67.24          & \textbf{63.61} & 59.68          & 67.24           & \textbf{63.61}  & 59.67           & 64.06          \\
           && ABAT-PGD 0.01  & 69.48          & 65.53          & 57.19          & 48.70          & 65.53          & 57.15          & 48.55          & 65.53           & 57.15           & 48.50           & 58.33          \\
           && ABAT-PGD 0.03  & 70.31          & \textbf{67.82} & 62.55          & 56.95          & \textbf{67.82} & 62.55          & 56.92          & \textbf{67.82}  & 62.55           & 56.91           & 63.22          \\
           && ABAT-PGD 0.05  & 69.18          & 67.38          & 63.52          & \textbf{59.79} & 67.38          & 63.51          & \textbf{59.79} & 67.38           & 63.51           & \textbf{59.78}  & \textbf{64.12} \\\midrule

\multirow{10}{*}{DeepCNN}
 & \multirow{3}{*}{\shortstack[c]{w/o \\ EA}}& BT   & 71.78 & 67.05 & 56.67 & 46.24 & 67.04 & 56.47 & 45.91 & 67.03 & 56.42 & 45.83 & 58.04 \\
 & &AT-FGSM 0.01& 70.16 & 67.50 & 61.75 & 55.87 & 67.49 & 61.69 & 55.70 & 67.48 & 61.67 & 55.65 & 62.50 \\
 & &AT-PGD 0.01& 70.92 & 68.15 & 62.35 & 56.34 & 68.14 & 62.30 & 56.19 & 68.14 & 62.29 & 56.13 & 63.10 \\ \cmidrule(r){2-14}
& \multirow{7}{*}{\shortstack[c]{with \\ EA}}& BT          & 71.77          & 64.81          & 49.39          & 35.28          & 64.75          & 48.85          & 33.76          & 64.73           & 48.71           & 33.21           & 51.53          \\
           && ABAT-FGSM 0.01 & 72.83          & 68.07          & 58.11          & 48.04          & 68.04          & 57.96          & 47.54          & 68.03           & 57.86           & 47.37           & 59.38          \\
           && ABAT-FGSM 0.03 & 73.34          & 70.32          & 64.02          & 57.47          & 70.31          & 63.96          & 57.24          & 70.31           & 63.92           & 57.17           & 64.81          \\
           && ABAT-FGSM 0.05 & 73.00          & \textbf{70.85} & 65.94          & 61.27          & \textbf{70.85} & 65.92          & 61.20          & \textbf{70.85}  & 65.90           & 61.18           & 66.70          \\
           && ABAT-PGD 0.01  & 72.78          & 68.24          & 58.26          & 48.12          & 68.24          & 58.11          & 47.55          & 68.23           & 58.06           & 47.43           & 59.50          \\
           && ABAT-PGD 0.03  & \textbf{73.61} & 70.54          & 64.42          & 57.90          & 70.54          & 64.38          & 57.74          & 70.54           & 64.38           & 57.69           & 65.17          \\
           && ABAT-PGD 0.05  & 72.70          & 70.76          & \textbf{66.02} & \textbf{61.41} & 70.76          & \textbf{65.99} & \textbf{61.34} & 70.76           & \textbf{65.99}  & \textbf{61.33}  & \textbf{66.71} \\\midrule

\multirow{10}{*}{ShallowCNN}
 & \multirow{3}{*}{\shortstack[c]{w/o \\ EA}}& BT    & 62.97 & 59.96 & 53.62 & 47.43 & 59.96 & 53.62 & 47.42 & 59.96 & 53.62 & 47.41 & 54.60 \\
 && AT-FGSM 0.01& 61.94 & 59.79 & 55.10 & 50.67 & 59.79 & 55.10 & 50.67 & 59.79 & 55.10 & 50.67 & 55.86 \\
 && AT-PGD 0.01& 62.32 & 60.17 & 55.47 & 51.20 & 60.17 & 55.46 & 51.19 & 60.17 & 55.46 & 51.18 & 56.28 \\ \cmidrule(r){2-14}
& \multirow{7}{*}{\shortstack[c]{with \\ EA}}& BT           & 66.61          & 60.77          & 49.50          & 37.35          & 60.77          & 49.46          & 37.31          & 60.77           & 49.46           & 37.29           & 50.93          \\
           && ABAT-FGSM 0.01 & 68.28          & 64.34          & 55.81          & 47.10          & 64.34          & 55.80          & 47.03          & 64.34           & 55.80           & 47.03           & 56.99          \\
           && ABAT-FGSM 0.03 & 68.66          & 66.04          & 60.95          & 55.33          & 66.04          & 60.94          & 55.30          & 66.04           & 60.94           & 55.30           & 61.55          \\
           && ABAT-FGSM 0.05 & 67.50          & 65.64          & \textbf{61.96} & 57.98          & 65.64          & \textbf{61.96} & 57.97          & 65.64           & \textbf{61.96}  & 57.97           & 62.42          \\
           && ABAT-PGD 0.01  & 68.44          & 64.38          & 55.94          & 47.20          & 64.38          & 55.93          & 47.17          & 64.38           & 55.93           & 47.15           & 57.09          \\
           && ABAT-PGD 0.03  & \textbf{68.81} & \textbf{66.17} & 61.05          & 55.48          & \textbf{66.17} & 61.04          & 55.46          & \textbf{66.17}  & 61.04           & 55.46           & 61.68          \\
           && ABAT-PGD 0.05  & 67.44          & 65.69          & 61.93          & \textbf{58.02} & 65.69          & 61.92          & \textbf{58.00} & 65.69           & 61.92           & \textbf{58.00}  & \textbf{62.43}\\
\bottomrule
\end{tabular}
\end{table*}

\begin{table*}[htbp] \centering \setlength{\tabcolsep}{2mm}\scriptsize
\caption{BCAs of different training approaches under benign samples and various attacks on ERP.}\label{tab:ERP_within}
\begin{tabular}{c|c|cccccccccccc} \toprule
\multirow{2}{*}{Model}&\multirow{2}{*}{EA}
&\multirow{2}{*}{Training}
&No&FGSM &FGSM &FGSM &PGD &PGD &PGD& AutoAttack& AutoAttack& AutoAttack&\multirow{2}{*}{\makebox[0.03\textwidth][c]{Avg.}}\\
&&&Attack&0.01&0.03&0.05&0.01&0.03&0.05&0.01&0.03&0.05 \\ \midrule
\multirow{16}{*}{EEGNet}
&\multirow{5}{*}{\shortstack[c]{w/o \\ EA}}& BT   & 84.97 & 77.15 & 57.04 & 36.96 & 77.12 & 56.46 & 35.12 & 77.11 & 56.37 & 34.78 & 59.31 \\
 & &AT-FGSM 0.01 & 85.87 & 81.61 & 71.24 & 58.73 & 81.60 & 71.11 & 58.24 & 81.60 & 71.08 & 58.17 & 71.93 \\
 & &AT-FGSM 0.03 & 83.77 & 81.49 & 76.06 & 69.90 & 81.49 & 76.05 & 69.79 & 81.49 & 76.04 & 69.78 & 76.59 \\
 & &AT-PGD 0.01& 85.73 & 81.71 & 71.30 & 58.82 & 81.71 & 71.18 & 58.42 & 81.71 & 71.15 & 58.32 & 72.00 \\
  & &AT-PGD 0.03& 83.66 & 81.29 & 75.91 & 69.80 & 81.28 & 75.88 & 69.72 & 81.28 & 75.88 & 69.72 & 76.44  \\ \cmidrule(r){2-14}
& \multirow{11}{*}{\shortstack[c]{with \\ EA}}&BT          & 84.37          & 81.41          & 74.49          & 67.03          & 81.41          & 74.45          & 66.92          & 81.41           & 74.45           & 66.88           & 75.28          \\
           && ABAT-FGSM 0.01 & 85.04          & 82.46          & 76.83          & 70.57          & 82.46          & 76.81          & 70.44          & 82.46           & 76.81           & 70.43           & 77.43          \\
           && ABAT-FGSM 0.03 & 85.95          & 83.95          & 79.72          & 74.76          & 83.95          & 79.72          & 74.71          & 83.95           & 79.72           & 74.71           & 80.11          \\

           && ABAT-FGSM 0.05 & 86.21          & 84.71 & 81.31 &77.70 & 84.71 & 81.30 & 77.65 & 84.71  & 81.30  & 77.65  & 81.73 \\
           && ABAT-FGSM 0.07 & \textbf{86.23} & \textbf{84.86} & 82.06 & 79.15 & \textbf{84.86} & 82.06 & 79.14 & \textbf{84.86} & 82.06 & 79.13 & 82.44         \\
           && ABAT-FGSM 0.09 & 85.82 & 84.82 &\textbf{82.46} &\textbf{79.93}& 84.82 & \textbf{82.46} &\textbf{79.93} & 84.82 & \textbf{82.46} & \textbf{79.93} & \textbf{82.75}\\

           && ABAT-PGD 0.01  & 85.19          & 82.97          & 77.07          & 70.74          & 82.97          & 77.06          & 70.63          & 82.97           & 77.06           & 70.63           & 77.73          \\
           && ABAT-PGD 0.03  & 86.07          & 84.13          & 79.75          & 74.90          & 84.13          & 79.74          & 74.76          & 84.13           & 79.74           & 74.76           & 80.21          \\
           && ABAT-PGD 0.05  & 86.22 & 84.61          & 81.14          & 77.35          & 84.61          & 81.12          & 77.32          & 84.61           & 81.12           & 77.31           & 81.54          \\
           && ABAT-PGD 0.07  & 86.15 & 84.71 & 82.08 & 79.24 & 84.71 & 82.07 & 79.21 & 84.71 & 82.07 & 79.21 & 82.41\\
           && ABAT-PGD 0.09  & 85.90 & 84.81 & 82.38 & 79.90 & 84.81 & 82.38 & 79.89 & 84.81 & 82.38 & 79.89 & 82.72         \\\midrule

\multirow{16}{*}{DeepCNN}
 & \multirow{5}{*}{\shortstack[c]{w/o \\ EA}}& BT   & 83.56 & 75.63 & 54.71 & 34.68 & 75.61 & 54.43 & 34.01 & 75.60 & 54.30 & 33.75 & 57.63 \\
 & &AT-FGSM 0.01 & 84.18 & 79.79 & 68.98 & 55.97 & 79.78 & 68.91 & 55.69 & 79.78 & 68.87 & 55.53 & 69.75 \\
  & &AT-FGSM 0.03 & 82.86 & 80.09 & 73.95 & 67.11 & 80.09 & 73.93 & 67.02 & 80.09 & 73.92 & 67.01 & 74.61\\
 & &AT-PGD 0.01 & 84.02 & 79.56 & 68.64 & 55.77 & 79.55 & 68.56 & 55.45 & 79.54 & 68.51 & 55.34 & 69.49 \\
  & &AT-PGD 0.03 &83.30 & 80.52 & 74.28 & 67.32 & 80.52 & 74.26 & 67.26 & 80.51 & 74.24 & 67.23 & 74.94\\ \cmidrule(r){2-14}
& \multirow{11}{*}{\shortstack[c]{with \\ EA}}&BT         & 82.43          & 78.66          & 70.90          & 62.28          & 78.64          & 70.61          & 61.48          & 78.64           & 70.52           & 61.12           & 71.53          \\
           & &ABAT-FGSM 0.01 & 83.10          & 80.18          & 73.37          & 66.33          & 80.16          & 73.28          & 65.92          & 80.16           & 73.24           & 65.79           & 74.15          \\
           & &ABAT-FGSM 0.03 & 83.70          & 81.44          & 76.10          & 70.05          & 81.44          & 76.00          & 69.83          & 81.44           & 75.98           & 69.75           & 76.57          \\
           & &ABAT-FGSM 0.05 & 84.38 & 82.57 & 78.06 & 72.83          & 82.57 & 78.01 & 72.66          & 82.57  & 77.99  & 72.61           & 78.42 \\
           & &ABAT-FGSM 0.07 & 84.82 & 83.20 & 79.59 & 75.01 & 83.19 & 79.58 & 74.93 & 83.19 & 79.57 & 74.88 & 79.80\\
           & &ABAT-FGSM 0.09 & 85.24 & 83.77 & 80.48 & 76.82 & 83.77 & 80.46 & 76.80 & 83.77 & 80.46 & 76.76 & 80.83 \\
           & &ABAT-PGD 0.01  & 82.73          & 79.76          & 73.28          & 66.37          & 79.73          & 73.09          & 65.94          & 79.71           & 73.05           & 65.82           & 73.95          \\
           & &ABAT-PGD 0.03  & 83.59          & 81.32          & 76.01          & 70.21          & 81.32          & 75.98          & 69.99          & 81.32           & 75.97           & 69.93           & 76.56          \\
           & &ABAT-PGD 0.05  & 84.16          & 82.21          & 77.90          & 73.01 & 82.21          & 77.87          & 72.92 & 82.21           & 77.87           & 72.88  & 78.32          \\
           & &ABAT-PGD 0.07  &84.67 & 83.07 & 79.38 & 75.03 & 83.07 & 79.36 & 74.97 & 83.07 & 79.35 & 74.95 & 79.69 \\
           & &ABAT-PGD 0.09  & \textbf{85.41} & \textbf{83.83} & \textbf{80.62} & \textbf{77.10} & \textbf{83.83} & \textbf{80.62} & \textbf{77.07} & \textbf{83.83} & \textbf{80.62} & \textbf{77.06} & \textbf{81.00}\\\midrule

\multirow{16}{*}{ShallowCNN}
 & \multirow{5}{*}{\shortstack[c]{w/o \\ EA}}&BT    & 84.54 & 76.49 & 55.63 & 35.65 & 76.48 & 55.43 & 35.07 & 76.46 & 55.41 & 34.98 & 58.61 \\
 & &AT-FGSM 0.01 & 84.85 & 80.73 & 70.69 & 58.58 & 80.73 & 70.65 & 58.39 & 80.73 & 70.64 & 58.36 & 71.44 \\
  && AT-FGSM 0.03 &83.31 & 80.97 & 76.01 & 69.51 & 80.97 & 76.00 & 69.49 & 80.97 & 76.00 & 69.48 & 76.27\\
 & &AT-PGD 0.01 & 84.92 & 80.91 & 70.67 & 58.45 & 80.91 & 70.64 & 58.30 & 80.91 & 70.62 & 58.27 & 71.46 \\
  & &AT-PGD 0.03  &83.15 & 80.89 & 75.85 & 69.35 & 80.89 & 75.85 & 69.33 & 80.89 & 75.85 & 69.33 & 76.14 \\ \cmidrule(r){2-14}
&\multirow{11}{*}{\shortstack[c]{with \\ EA}}& BT         & 82.53          & 79.59          & 73.57          & 65.88          & 79.59          & 73.56          & 65.86          & 79.59           & 73.56           & 65.85           & 73.96          \\
           & &ABAT-FGSM 0.01 & 83.34          & 80.97          & 75.61          & 69.09          & 80.97          & 75.61          & 69.06          & 80.97           & 75.61           & 69.06           & 76.03          \\
           & &ABAT-FGSM 0.03 & 84.65          & 82.59          & 78.16          & 72.77          & 82.59          & 78.16          & 72.76          & 82.59           & 78.16           & 72.76           & 78.52          \\
           & &ABAT-FGSM 0.05 & 85.11          & 83.26          & 79.26          & 75.12          & 83.26          & 79.25          & 75.11          & 83.26           & 79.25           & 75.10           & 79.80          \\
           & &ABAT-FGSM 0.07 & 85.45 & 83.97 & 80.40 & 76.81 & 83.97 & 80.40 & 76.81 & 83.97 & 80.40 & 76.81 & 80.90\\
           & &ABAT-FGSM 0.09 & \textbf{85.62} & 84.17 & 81.28 & \textbf{78.04} & 84.17 & 81.28 & \textbf{78.03} & 84.17 & 81.28 & \textbf{78.03} & \textbf{81.61}\\
           & &ABAT-PGD 0.01  & 83.39          & 81.11          & 75.63          & 69.11          & 81.11          & 75.63          & 69.11          & 81.11           & 75.63           & 69.11           & 76.09          \\
           & &ABAT-PGD 0.03  & 84.57          & 82.50          & 78.04          & 72.78          & 82.50          & 78.04          & 72.77          & 82.50           & 78.04           & 72.77           & 78.45          \\
           & &ABAT-PGD 0.05  & 85.19 & 83.32 & 79.49 & 75.32 & 83.32 & 79.49 & 75.32 & 83.32  & 79.49  & 75.32  & 79.96 \\
           & &ABAT-PGD 0.07  & 85.36 & 83.93 & 80.52 & 77.03 & 83.93 & 80.52 & 77.02 & 83.93 & 80.52 & 77.02 & 80.98\\
           & &ABAT-PGD 0.09  & 85.54 & \textbf{84.22} & \textbf{81.30} & 77.95 & \textbf{84.22} & \textbf{81.30} & 77.93 & \textbf{84.22} & \textbf{81.30} & 77.93 & 81.59\\
\bottomrule
\end{tabular}
\end{table*}

\subsection{Online Cross-Session Performance on Benign Samples}

\begin{figure}\centering
\subfigure[]{\includegraphics[width=.7\linewidth,clip]{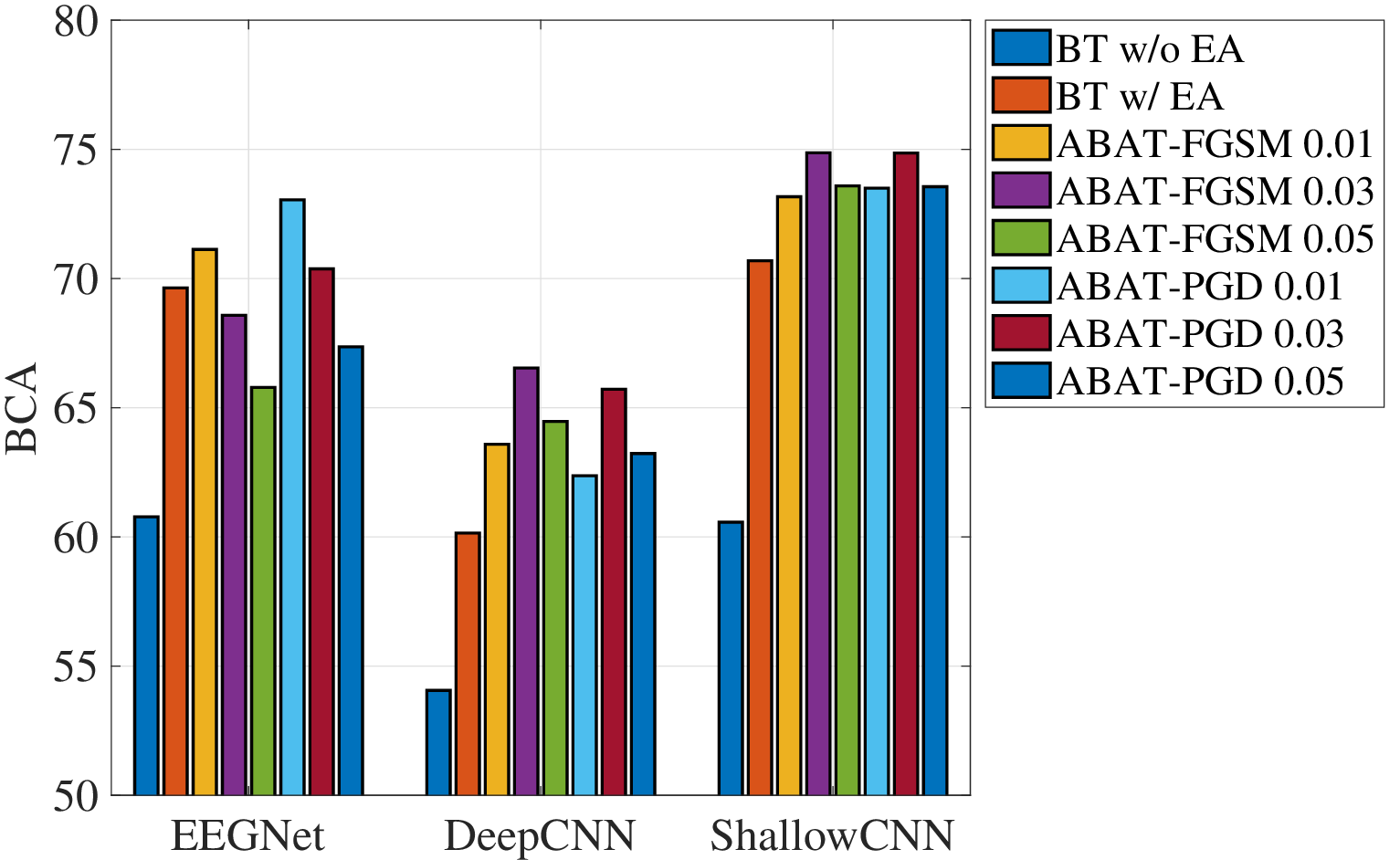}\label{fig:oa}}
\subfigure[]{\includegraphics[width=.7\linewidth,clip]{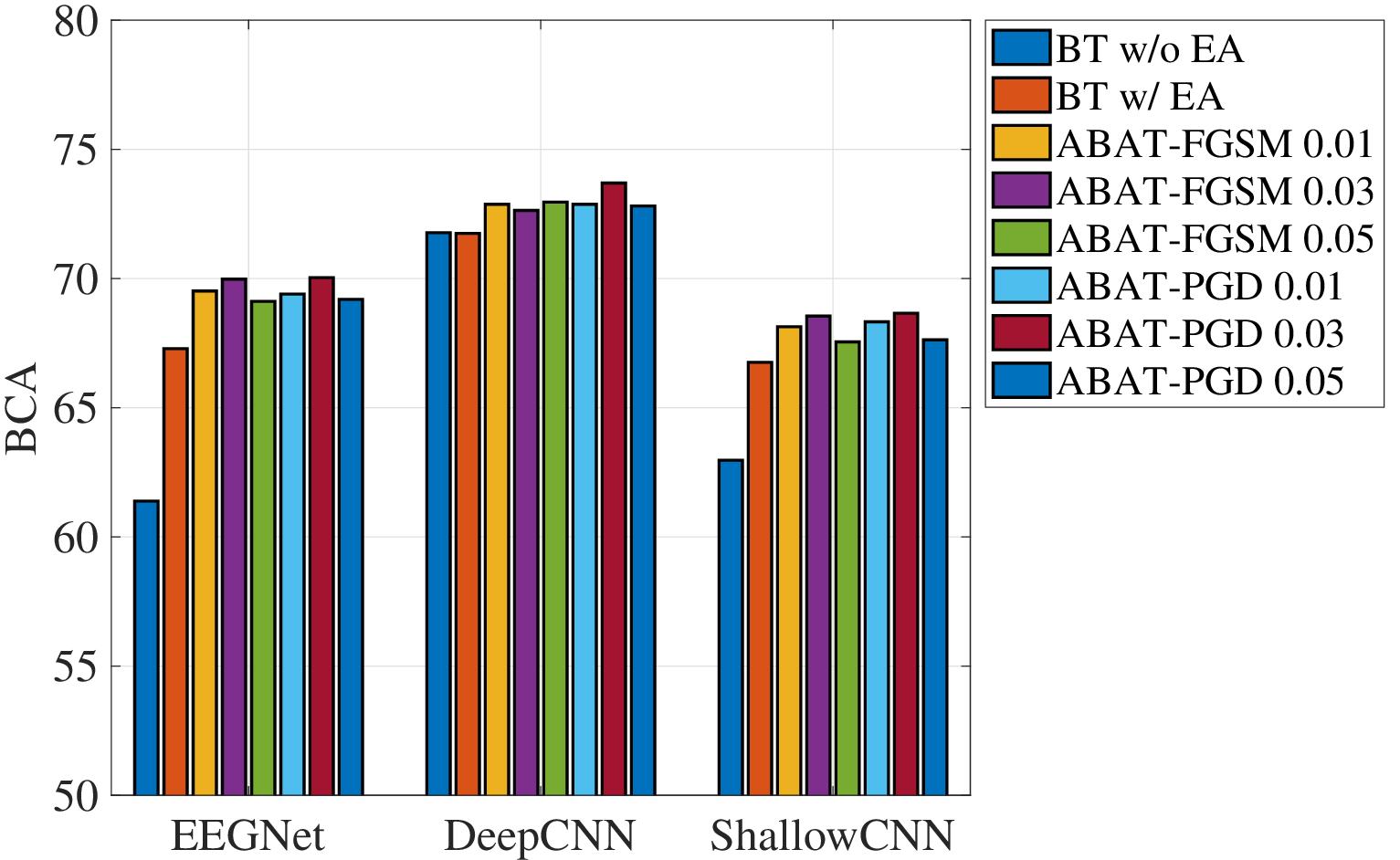}\label{fig:ob}}
\caption{Online cross-session performance of ABAT using incremental EA on (a) MI4 and (b) P300.}\label{online}
\end{figure}

We simulated the online cross-session EEG classification scenario and performed incremental EA for each incoming EEG trail in the target session on MI4 and P300 datasets. The classifiers were the same as those in Subsection~\ref{sect:ma}. The results are shown in Fig.~\ref{online}. When EA was not used, the results for online scenarios were the same as those for offline scenarios.

Using incremental EA in online scenarios improved model performance, and ABAT can further improve the BCAs.

\subsection{ABAT Using Pre-Trained Models}

In real-world applications, we could pre-train a classifier on other subjects' EEG data, and then fine-tune it with target subject' data. The results on MI4 and MI6 are shown in Tables~\ref{tab:MI4_pretrain}-\ref{tab:MI6_pretrain}, respectively.

Compared with BCAs without pre-training in Tables~\ref{tab:MI4_within}-\ref{tab:MI6_within}, when EA was used, pre-training achieved higher BCAs on benign samples of the target user; however, the BCAs decreased on adversarial examples with larger perturbations. Nevertheless, ABAT can further improve the BCAs on both benign samples and adversarial examples.

\begin{table*}[htbp] \centering \setlength{\tabcolsep}{2mm} \scriptsize
\caption{BCAs of different training approaches on pre-trained models under various attacks on MI4.}\label{tab:MI4_pretrain}
\begin{tabular}{c|c|cccccccccccc} \toprule
\multirow{2}{*}{Model} &\multirow{2}{*}{EA}
&\multirow{2}{*}{Training}
&No&FGSM &FGSM &FGSM &PGD &PGD &PGD& AutoAttack& AutoAttack& AutoAttack&\multirow{2}{*}{\makebox[0.03\textwidth][c]{Avg.}}\\
&&&Attack&0.01&0.03&0.05&0.01&0.03&0.05&0.01&0.03&0.05 \\ \midrule
\multirow{10}{*}{EEGNet}
 & \multirow{3}{*}{\shortstack[c]{w/o \\ EA}}& BT                 & 67.40 & 31.38 & 5.27 & 1.36 & 30.74 & 3.69 & 0.40 & 30.25 & 3.41 & 0.30 & 17.42 \\
 & &AT-FGSM 0.01 &61.30 & 51.66 & 34.43 & 19.97 & 51.65 & 33.81 & 18.80 & 51.52 & 33.36 & 18.29 & 37.48  \\
 & &AT-PGD 0.01      & 59.89 & 50.69 & 33.35 & 19.28 & 50.76 & 32.97 & 18.08 & 50.59 & 32.66 & 17.59 & 36.59 \\ \cmidrule(r){2-14}
& \multirow{7}{*}{\shortstack[c]{with \\ EA}}& BT           & 74.31          & 61.09          & 37.56          & 20.72          & 61.03          & 36.81          & 19.69          & 60.97           & 36.54           & 18.97           & 42.77          \\
           & &ABAT-FGSM 0.01 & 75.75          & 67.68          & 50.57          & 35.79          & 67.68          & 50.24          & 34.97          & 67.67           & 49.96           & 34.52           & 53.48          \\
           & &ABAT-FGSM 0.03 & 74.14          & 70.25          & 60.30          & 49.87          & 70.25          & 60.17          & 49.54          & 70.25           & 60.10           & 49.24           & 61.41          \\
           & &ABAT-FGSM 0.05 & 70.63          & 67.75          & 61.57          & \textbf{54.37} & 67.75          & 61.60          & \textbf{54.27} & 67.75           & 61.52           & \textbf{54.03}  & 62.12          \\
           & &ABAT-PGD 0.01  & \textbf{76.43} & 68.25          & 51.56          & 36.28          & 68.24          & 51.27          & 35.58          & 68.22           & 51.04           & 35.20           & 54.21          \\
           & &ABAT-PGD 0.03  & 74.67          & \textbf{70.32} & 60.70          & 50.24          & \textbf{70.33} & 60.64          & 49.96          & \textbf{70.31}  & 60.58           & 49.76           & 61.75          \\
           & &ABAT-PGD 0.05  & 70.86          & 67.68          & \textbf{61.92} & 54.36          & 67.67          & \textbf{61.95} & 54.17          & 67.67           & \textbf{61.87}  & 53.99           & \textbf{62.21} \\\midrule

\multirow{10}{*}{DeepCNN}
 & \multirow{3}{*}{\shortstack[c]{w/o \\ EA}}& BT                  & 61.29 & 24.02 & 1.29 & 0.06 & 23.35 & 0.99 & 0.03 & 23.03 & 0.82 & 0.03 & 13.49 \\
 & &AT-FGSM 0.01& 55.31 & 44.66 & 25.95 & 13.30 & 44.60 & 25.41 & 12.37 & 44.53 & 25.14 & 12.06 & 30.33 \\
 & &AT-PGD 0.01      & 55.52 & 44.88 & 26.43 & 13.41 & 44.77 & 25.77 & 12.73 & 44.73 & 25.51 & 12.26 & 30.60 \\ \cmidrule(r){2-14}
& \multirow{7}{*}{\shortstack[c]{with \\ EA}}&BT        & 71.66          & 48.01          & 17.80          & 6.73           & 46.95          & 14.26          & 3.24           & 46.71           & 12.82           & 2.17            & 27.03          \\
           & &ABAT-FGSM 0.01 & \textbf{73.68} & 61.54          & 38.53          & 20.42          & 61.38          & 37.31          & 18.65          & 61.32           & 36.74           & 17.57           & 42.71          \\
           & &ABAT-FGSM 0.03 & 72.70          & \textbf{64.87} & 48.26          & 32.84          & \textbf{64.85} & 47.90          & 32.02          & \textbf{64.84}  & 47.75           & 31.56           & 50.76          \\
           & &ABAT-FGSM 0.05 & 68.99          & 63.44          & 51.89          & 40.24          & 63.44          & 51.71          & 40.01          & 63.44           & 51.63           & 39.78           & 53.46          \\
           & &ABAT-PGD 0.01  & 72.12          & 59.63          & 37.26          & 19.91          & 59.54          & 36.03          & 18.33          & 59.50           & 35.49           & 17.32           & 41.51          \\
           & &ABAT-PGD 0.03  & 72.61          & 64.51          & 47.60          & 33.27          & 64.45          & 47.31          & 32.47          & 64.45           & 47.02           & 31.85           & 50.55          \\
           & &ABAT-PGD 0.05  & 69.25          & 64.06          & \textbf{51.90} & \textbf{40.88} & 64.06          & \textbf{51.77} & \textbf{40.46} & 64.06           & \textbf{51.67}  & \textbf{40.23}  & \textbf{53.83} \\ \midrule

\multirow{10}{*}{ShallowCNN}
 & \multirow{3}{*}{\shortstack[c]{w/o \\ EA}}& BT               & 65.63 & 32.30 & 5.05 & 0.71 & 31.97 & 4.17 & 0.35 & 31.75 & 3.69 & 0.26 & 17.59\\
 & &AT-FGSM 0.01& 60.62 & 48.62 & 28.07 & 14.16 & 48.61 & 27.75 & 13.43 & 48.52 & 27.38 & 12.87 & 33.00 \\
 & &AT-PGD 0.01      & 60.69 & 49.10 & 27.89 & 14.29 & 49.07 & 27.44 & 13.55 & 49.02 & 27.20 & 12.86 & 33.11 \\ \cmidrule(r){2-14}
& \multirow{7}{*}{\shortstack[c]{with \\ EA}}& BT         & 76.63          & 62.05          & 35.61          & 18.31          & 61.90          & 34.80          & 16.90          & 61.82           & 34.32           & 16.35           & 41.87          \\
           && ABAT-FGSM 0.01 & \textbf{77.56} & 68.78          & 49.37          & 33.51          & 68.69          & 49.00          & 32.81          & 68.66           & 48.84           & 32.46           & 52.97          \\
           && ABAT-FGSM 0.03 & 77.06          & \textbf{70.15} & 55.03          & 40.77          & \textbf{70.13} & 54.77          & 40.38          & \textbf{70.13}  & 54.71           & 40.16           & 57.33          \\
           && ABAT-FGSM 0.05 & 75.81          & 70.00          & \textbf{57.50} & 44.05          & 69.98          & \textbf{57.38} & 43.83          & 69.98           & \textbf{57.30}  & 43.60           & \textbf{58.94} \\
           && ABAT-PGD 0.01  & 76.81          & 68.09          & 49.24          & 33.27          & 68.07          & 49.00          & 32.64          & 68.03           & 48.84           & 32.20           & 52.62          \\
           && ABAT-PGD 0.03  & 76.62          & 70.13          & 55.04          & 41.27          & 70.10          & 54.86          & 40.88          & 70.11           & 54.73           & 40.73           & 57.45          \\
           && ABAT-PGD 0.05  & 75.64          & 69.68          & 57.12          & \textbf{44.24} & 69.68          & 56.98          & \textbf{44.06} & 69.68           & 56.97           & \textbf{43.87}  & 58.79   \\
\bottomrule
\end{tabular}
\end{table*}

\begin{table*}[htbp] \centering \setlength{\tabcolsep}{2mm}
\scriptsize
\caption{BCAs of different training approaches on pre-trained models under various attacks on MI6.}\label{tab:MI6_pretrain}
\begin{tabular}{c|c|cccccccccccc} \toprule
\multirow{2}{*}{Model}&\multirow{2}{*}{EA}
&\multirow{2}{*}{Training}
&No&FGSM &FGSM &FGSM &PGD &PGD &PGD& AutoAttack& AutoAttack& AutoAttack&\multirow{2}{*}{\makebox[0.03\textwidth][c]{Avg.}}\\
&&&Attack&0.01&0.03&0.05&0.01&0.03&0.05&0.01&0.03&0.05 \\ \midrule
\multirow{10}{*}{EEGNet}
 & \multirow{3}{*}{\shortstack[c]{w/o \\ EA}}&BT                 & 49.31 & 17.77 & 1.51 & 0.41 & 17.22 & 1.08 & 0.14 & 16.67 & 0.87 & 0.12 & 10.51 \\
 & &AT-FGSM 0.01& 49.58 & 38.26 & 18.54 & 7.97 & 38.21 & 17.83 & 6.84 & 38.13 & 17.27 & 6.40 & 23.90 \\
 & &AT-PGD 0.01      & 48.75 & 37.73 & 18.46 & 7.98 & 37.73 & 17.84 & 7.07 & 37.62 & 17.41 & 6.59 & 23.72 \\ \cmidrule(r){2-14}
& \multirow{7}{*}{\shortstack[c]{with \\ EA}}& BT         & 48.97          & 23.38          & 4.01           & 0.56           & 22.94          & 3.16           & 0.34           & 22.74           & 2.78            & 0.26            & 12.91          \\
           & &ABAT-FGSM 0.01 & 57.35          & 41.93          & 19.02          & 7.29           & 41.86          & 17.96          & 6.10           & 41.80           & 17.43           & 5.49            & 25.62          \\
           & &ABAT-FGSM 0.03 & \textbf{61.94} & 55.17          & 42.08          & 29.67          & 55.15          & 41.87          & 28.94          & 55.11           & 41.58           & 28.49           & 44.00          \\
           & &ABAT-FGSM 0.05 & 60.33          & \textbf{56.26} & 47.58          & 38.97          & \textbf{56.27} & 47.51          & 38.95          & \textbf{56.25}  & 47.37           & 38.64           & 48.81          \\
           & &ABAT-PGD 0.01  & 57.39          & 41.73          & 18.05          & 6.96           & 41.62          & 17.17          & 5.72           & 41.58           & 16.66           & 5.25            & 25.21          \\
           & &ABAT-PGD 0.03  & 61.24          & 54.95          & 42.00          & 30.16          & 54.92          & 41.80          & 29.56          & 54.91           & 41.64           & 29.18           & 44.04          \\
           & &ABAT-PGD 0.05  & 60.21          & 56.04          & \textbf{47.77} & \textbf{39.44} & 56.04          & \textbf{47.68} & \textbf{39.26} & 56.03           & \textbf{47.57}  & \textbf{39.05}  & \textbf{48.91} \\\midrule

\multirow{10}{*}{DeepCNN}
 & \multirow{3}{*}{\shortstack[c]{w/o \\ EA}}&BT                 & 33.82 & 14.52 & 1.79 & 0.39 & 14.24 & 1.58 & 0.30 & 14.07 & 1.47 & 0.29 & 8.25 \\
 & &AT-FGSM 0.01& 30.98 & 22.14 & 9.54 & 3.58 & 22.08 & 9.29 & 3.05 & 22.06 & 9.11 & 2.87 & 13.47 \\
 & &AT-PGD 0.01      & 30.73 & 22.19 & 9.81 & 3.91 & 22.14 & 9.50 & 3.53 & 22.07 & 9.30 & 3.26 & 13.64 \\ \cmidrule(r){2-14}
 & \multirow{7}{*}{\shortstack[c]{with \\ EA}}& BT         & 49.55          & 18.69          & 1.69           & 0.21           & 17.78          & 0.97           & 0.06           & 17.16           & 0.78            & 0.04            & 10.69          \\
           && ABAT-FGSM 0.01 & 49.55          & 18.69          & 1.69           & 0.21           & 17.78          & 0.97           & 0.06           & 17.16           & 0.78            & 0.04            & 10.69          \\
           && ABAT-FGSM 0.03 & \textbf{53.29} & 41.80          & 23.19          & 11.43          & 41.72          & 22.55          & 10.49          & 41.69           & 22.18           & 9.90            & 27.82          \\
           && ABAT-FGSM 0.05 & 50.93          & 43.93          & 31.13          & 20.44          & 43.90          & 30.83          & 19.88          & 43.89           & 30.71           & 19.52           & 33.52          \\
           && ABAT-PGD 0.01  & 52.60          & 32.90          & 9.65           & 2.33           & 32.57          & 8.62           & 1.60           & 32.45           & 8.06            & 1.37            & 18.21          \\
           && ABAT-PGD 0.03  & 52.56          & 41.53          & 23.26          & 11.45          & 41.49          & 22.61          & 10.52          & 41.46           & 22.36           & 9.90            & 27.71          \\
           && ABAT-PGD 0.05  & 50.96          & \textbf{44.12} & \textbf{31.20} & \textbf{20.56} & \textbf{44.10} & \textbf{30.85} & \textbf{20.00} & \textbf{44.09}  & \textbf{30.78}  & \textbf{19.71}  & \textbf{33.64} \\\midrule

\multirow{10}{*}{ShallowCNN}
 & \multirow{3}{*}{\shortstack[c]{w/o \\ EA}}& BT                & 46.36 & 24.45 & 4.09 & 0.58 & 24.31 & 3.66 & 0.42 & 24.14 & 3.34 & 0.34 & 13.17 \\
 & &AT-FGSM 0.01& 43.40 & 32.58 & 15.62 & 5.92 & 32.56 & 15.26 & 5.47 & 32.52 & 15.01 & 5.14 & 20.35 \\
 & &AT-PGD 0.01      & 43.39 & 32.82 & 15.64 & 6.02 & 32.75 & 15.22 & 5.42 & 32.72 & 14.90 & 5.18 & 20.41\\ \cmidrule(r){2-14}
& \multirow{7}{*}{\shortstack[c]{with \\ EA}}& BT          & 57.25          & 34.36          & 8.49           & 1.93           & 33.93          & 6.54           & 0.67           & 33.77           & 5.42            & 0.42            & 18.28          \\
           && ABAT-FGSM 0.01 & 58.44          & 43.71          & 21.77          & 8.51           & 43.63          & 20.89          & 7.07           & 43.58           & 20.34           & 5.87            & 27.38          \\
           && ABAT-FGSM 0.03 & \textbf{59.77} & 48.92          & 30.44          & 16.97          & 48.88          & 30.03          & 16.20          & 48.84           & 29.82           & 15.64           & 34.55          \\
           && ABAT-FGSM 0.05 & 59.35          & \textbf{50.43} & \textbf{34.52} & 21.67          & \textbf{50.37} & \textbf{34.31} & 21.12          & \textbf{50.36}  & \textbf{34.19}  & 20.78           & \textbf{37.71} \\
           && ABAT-PGD 0.01  & 58.47          & 43.81          & 21.64          & 8.25           & 43.71          & 20.72          & 6.90           & 43.67           & 19.88           & 5.85            & 27.29          \\
           && ABAT-PGD 0.03  & 58.86          & 48.70          & 30.51          & 17.15          & 48.64          & 30.25          & 16.46          & 48.62           & 30.11           & 16.01           & 34.53          \\
           && ABAT-PGD 0.05  & 58.58          & 49.95          & 34.24          & \textbf{21.91} & 49.93          & 34.08          & \textbf{21.43} & 49.92           & 33.98           & \textbf{21.23}  & 37.52          \\
\bottomrule
\end{tabular}
\end{table*}

\subsection{Discussions}

\begin{figure*}\centering
\subfigure[]{\includegraphics[height=1.8in]{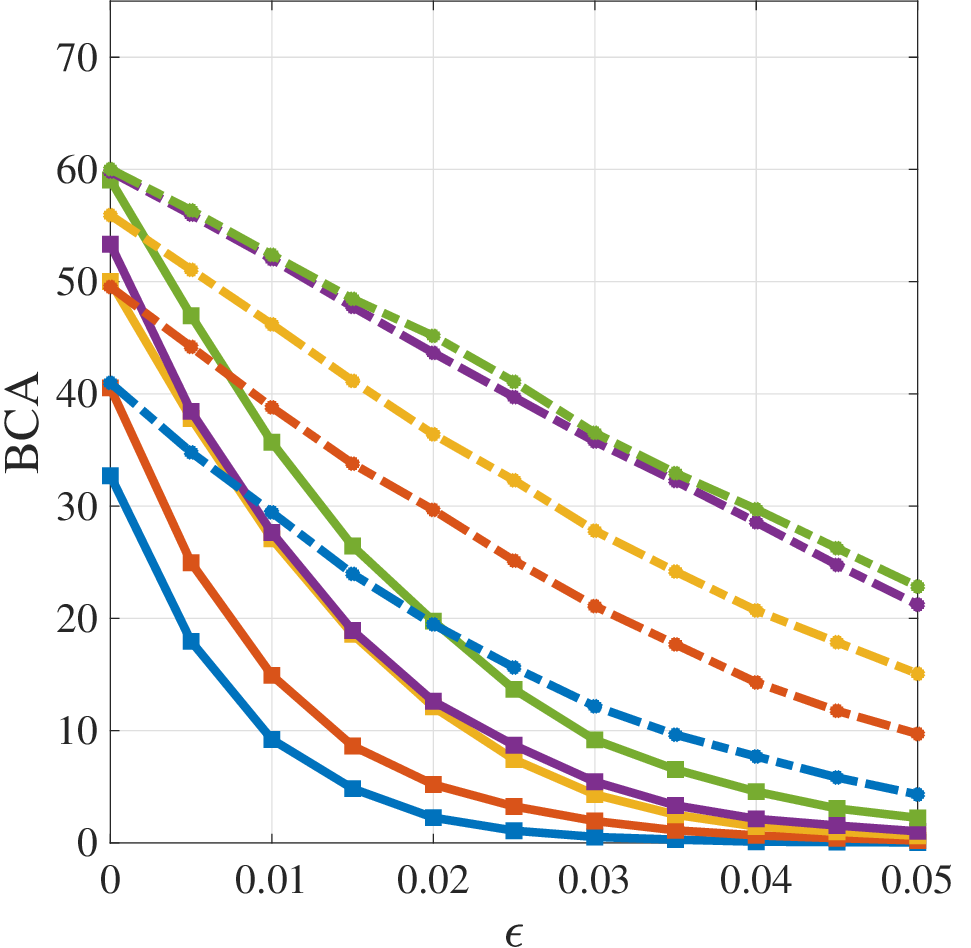}\label{fig:7a}}
\subfigure[]{\includegraphics[height=1.8in]{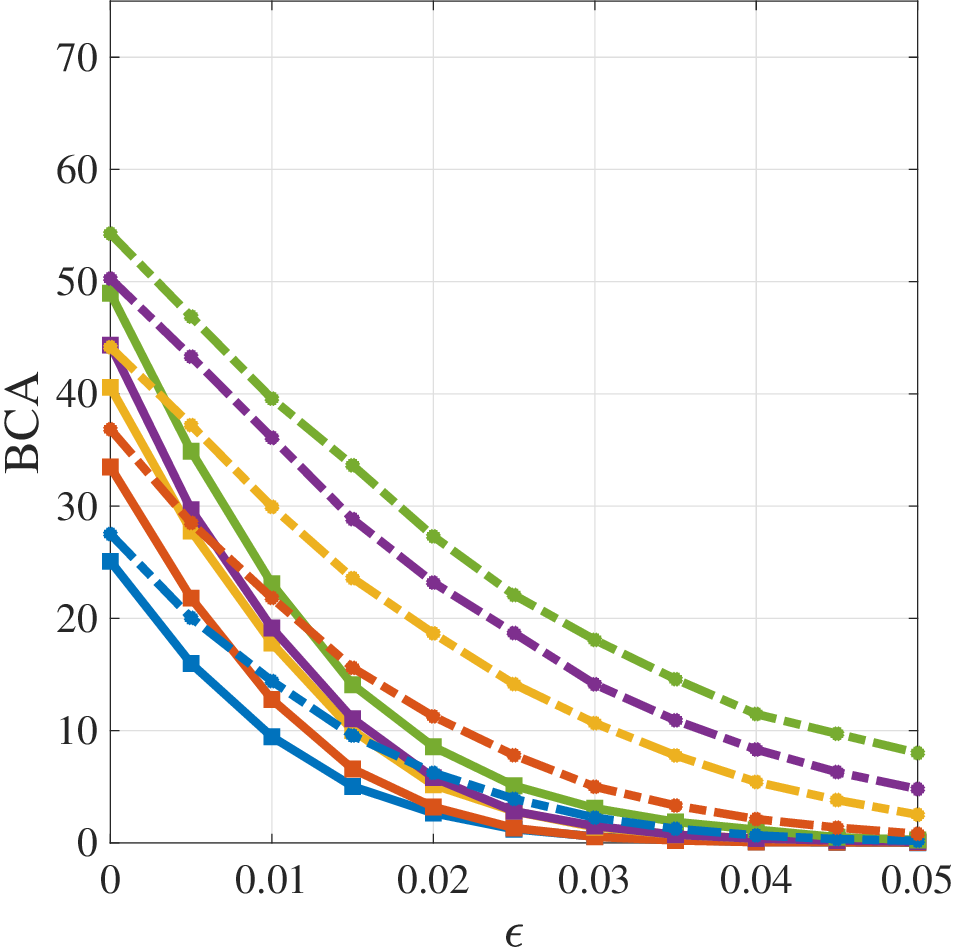}\label{fig:7b}}
\subfigure[]{\includegraphics[height=1.8in]{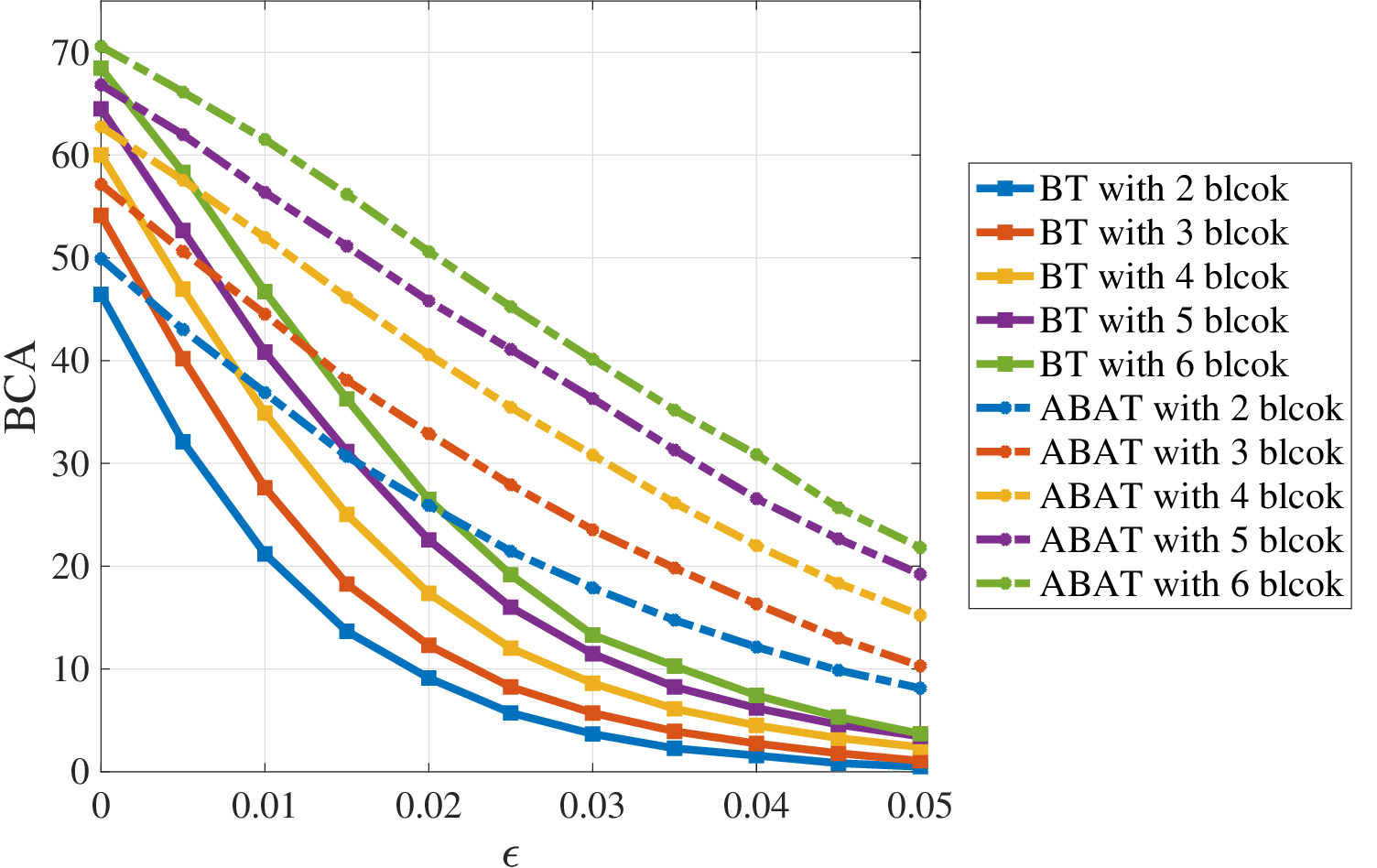}\label{fig:7c}}
\caption{BCAs of (a) EEGNet (b) DeepCNN and (c) ShallowCNN with different training data size and perturbation magnitude on MI6.}\label{ba}
\end{figure*}

To explore the influence of training data size on the classification performance in offline within-subject classification, we trained the classifiers using different number of blocks of data on MI6, and computed their BCAs under different AutoAttack amplitudes. We used EA in BT, and ABAT used PGD with $\epsilon=0.01$. The results are shown in Figs.~\ref{fig:7a}-\ref{fig:7c}. Perturbation amplitude 0 means benign samples.

It can be observed that increasing the training data size steadily improved the classification performance on benign samples, which is intuitive. However, the classifiers were still vulnerable to adversarial attacks. ABAT improved both the classification performance on benign samples with different training data sizes, and the robustness of the classifiers.


Subsection~\ref{sect:ma} pointed out that the optimal ABAT perturbation amplitude for a classifier to achieve their highest BCAs on benign samples may be positively correlated with its capacity. This subsection performs further investigations.

\begin{figure}\centering
\subfigure[]{\includegraphics[width=.5\linewidth,clip]{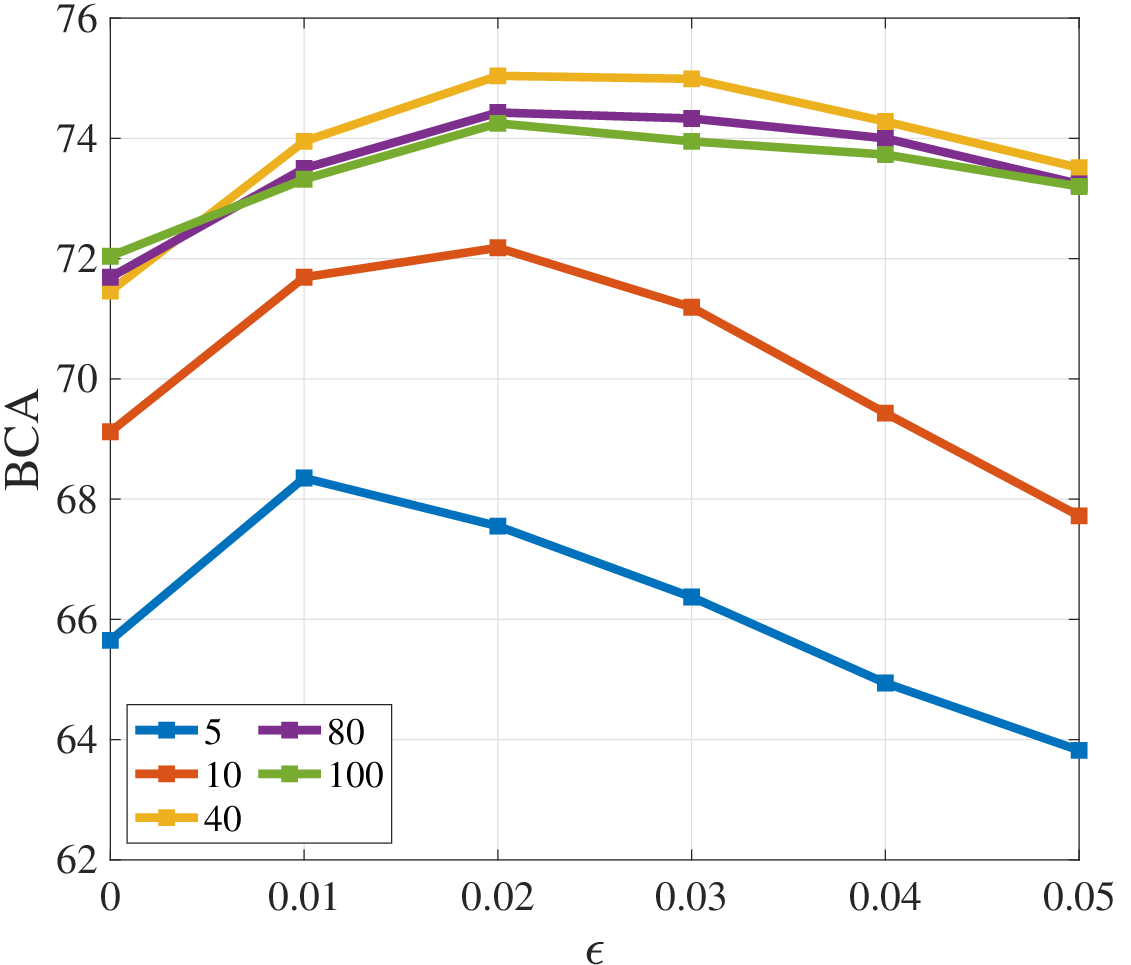}}
\subfigure[]{\includegraphics[width=.48\linewidth,clip]{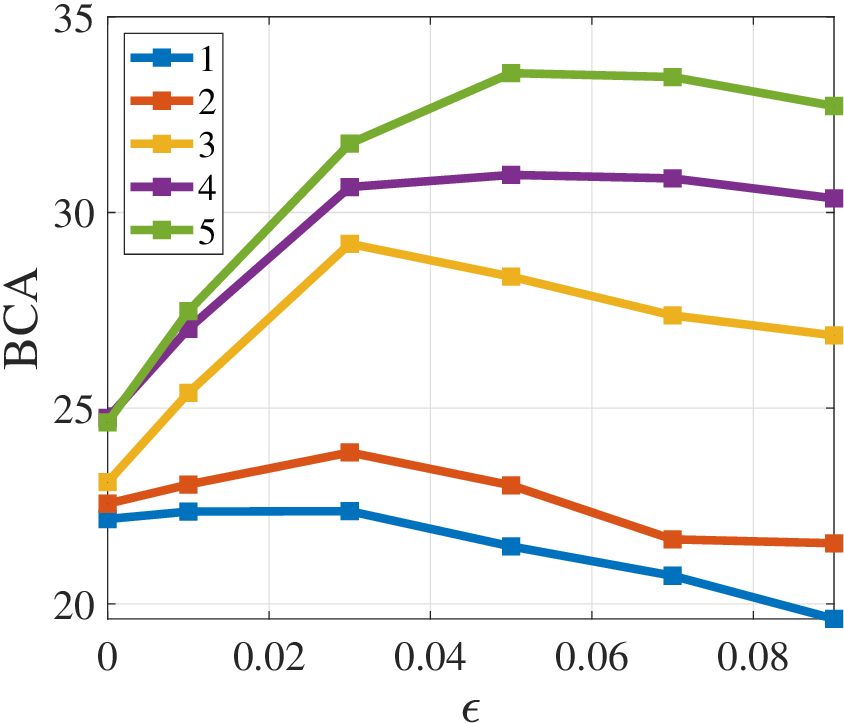}}
\caption{BCAs of (a) ShallowCNN on MI4, and (b) DeepCNN on MI6, for benign samples with different model capacities and perturbation magnitude.}\label{capa}
\end{figure}

We tested the BCAs of ShallowCNN with $\{5, 10, 40, 80, 100\}$ convolution kernels on MI4, and DeepCNN with $\{1-2-2, 2-4-8, 5-10-20, 10-20-40, 20-40-80\}$ convolution kernels (denoted as 1, 2, 3, 4 and 5, respectively) on MI6, for benign samples under different ABAT-PGD perturbation amplitudes. The results are shown in Fig.~\ref{capa}. Generally, as the number of model parameters increased, the optimal ABAT amplitude for benign samples also increased.

\section{Conclusions and Future Research} \label{sect:CFR}

This papers has proposed a simple yet effective ABAT approach to perform AT on aligned EEG data in order to make the trained model simultaneously more accurate and more robust. Experiments on four EEG datasets from two different BCI paradigms (MI and ERP), three CNN classifiers (EEGNet, ShallowCNN and DeepCNN) and three different experimental settings (offline within-subject cross-block/-session classification, online cross-session classification, and pre-trained classifiers) demonstrated its effectiveness. It is very intriguing that adversarial attacks, which are usually used to damage BCI systems, can be used in adversarial training to simultaneously improve the model accuracy and robustness.

Our future research will:
\begin{enumerate}
	\item Study how to perform ABAT for traditional EEG classifiers. This paper proposed ABAT for deep neural network EEG classifiers, but there are also many promising traditional classifiers \cite{Sadiq2019, Akbari2021,Sadiq2022,Akbari2023}, and it is useful to adapt ABAT to them.
	\item Study how ABAT can be used in cross-subject application, to increase the accuracy and robustness simultaneously. EEG data exhibit large individual differences \cite{drwuMITLBCI2022}, so this problem is very challenging.
	\item  Study how data preprocessing/denoising approaches, e.g., multiscale principal component analysis \cite{Sadiq2019, Sadiq2020,Sadiq2022a,Sadiq2022}, can be integrated with ABAT for even better performance.
\end{enumerate}



\end{document}